\documentclass{aa}
\usepackage{graphicx}
\usepackage{txfonts}
\usepackage{subcaption}
\usepackage{float}
\usepackage{color}
\usepackage{multirow}
\usepackage{soul}
\usepackage{hyperref}
\usepackage[dvipsnames]{xcolor}

\usepackage{pifont}
\newcommand{\cmark}{\ding{51}}
\newcommand{\xmark}{\ding{55}}
\usepackage{xcolor}
\newcommand{\ignore}[1]{}

\begin{document} 

   \title{Nuclear regions as seen with LOFAR international baselines}

   \subtitle{A high-resolution study of the recurrent activity}

   \author{N. Jurlin\inst{1}\fnmsep\inst{2}\fnmsep\thanks{Moved to Department of Astronomy, The University of Texas at Austin, 2515 Speedway, Stop C1400, Austin, TX 78712-1205, USA}\fnmsep\thanks{jurlin@utexas.edu},
          R. Morganti\inst{1}\fnmsep\inst{2},
          F. Sweijen\inst{3},
          L. K. Morabito\inst{4}\fnmsep\inst{5},
          M. Brienza\inst{6}\fnmsep\inst{7}\fnmsep\inst{8},
          P. Barthel\inst{1},
          G. K. Miley\inst{3}
          }

   \institute{Kapteyn Astronomical Institute, University of Groningen, PO Box 800, 9700 AV, Groningen, The Netherlands
    \and ASTRON, Netherlands Institute for Radio Astronomy, Oude Hoogeveensedijk 4, 7991 PD, Dwingeloo, The Netherlands
    \and Sterrewacht Leiden, University of Leiden, 2300 RA, Leiden, The Netherlands
    \and Centre for Extragalactic Astronomy, Department of Physics, Durham University, DH1 3LE, UK
    \and Institute for Computational Cosmology, Department of Physics, Durham University, DH1 3LE, UK
    \and INAF – Osservatorio di Astrofisica e Scienza dello Spazio di Bologna, Via P. Gobetti 93/3, 40129 Bologna, Italy
    \and Dipartimento di Fisica e Astronomia, Università di Bologna, Via P. Gobetti 93/2, I-40129 Bologna, Italy
    \and INAF---Istituto di Radio Astronomia, Via P. Gobetti 101, I-40129 Bologna, Italy}   


  \abstract
   {Radio galaxies dominate the sky at radio wavelengths and represent an essential piece in the galaxy evolution puzzle. High-resolution studies focused on statistical samples of radio galaxies are expected to shed light on the triggering mechanisms of the active galactic nucleus in their centre, alternating between the phases of activity and quiescence.}
   {In this work, we zoom in on the sub-arcsec radio structures in the central regions of the 35 radio galaxies in the area covering 6.6 $\rm deg^{2}$ of the Lockman Hole region. The sources studied here were previously classified as active, remnant, and candidate restarted radio galaxies based on the Low Frequency Array (LOFAR) observations at 150 MHz. We examine the morphologies and study the spectral properties of their central regions to explore their evolutionary stages and revise the morphological and spectral criteria used to select the initial sample.}
   {We use the newly available LOFAR 150 MHz image obtained using international baselines, yielding a resolution of 0.38$^{\prime\prime}$ $\times$ 0.30$^{\prime\prime}$, making this the first systematic study of the nuclear regions at such a high resolution and low frequency. We use publicly available images from the Faint Images of the Radio Sky at Twenty-cm survey at 1.4 GHz and the \textit{Karl G. Jansky} Very Large Array (VLA) Sky Survey at 3 GHz to achieve our goals.
   In addition, for one of the restarted candidates we present new dedicated observations with the VLA at 3 GHz.}
   {We have characterised the central regions of the radio galaxies in our sample and found various morphologies, some even mimicking well-known double-double radio galaxies but on a smaller scale, i.e. a few tens of kiloparsecs for the size of the restarted activity. We also see the beginnings of active jets or distinct detections unrelated to the large-scale structure. Furthermore, we have found a variety of radio spectra characterising the sources in our sample, i.e. flat, steep, or peaked in the frequency range between 150 and 3 GHz, indicative of the different life-cycle phases of the sources in our sample. 
   Based on these analyses, we confirm five out of six previously considered restarted candidates and identify three more restarted candidates from the active sample. As the number of restarted candidates still exceeds that of remnant candidates, this is consistent with previous results suggesting that the restarted phase can occur after a relatively short remnant phase (i.e. a few tens of millions of years).}
   {}

    \authorrunning{Jurlin et al.}
    \titlerunning{Nuclear regions of radio galaxies}
    \maketitle
%

\section{Introduction} \label{sec:introduction}
Feedback from radio jets associated with active galactic nuclei (AGN) plays a prominent role in the evolution of galaxies. However, the impact of radio jets on the host galaxy and surrounding medium is still not well understood. The fraction of radio-loud (jetted) AGN can change substantially based on radio power of the source and the stellar mass of the host galaxy, with a fraction of >30\% for massive ($\rm M_{*} = 5 \times 10^{11} M_{\sun}$) early-type galaxies with intermediate radio luminosities ($\rm L_{1.4~GHz} > 10^{24} W~Hz^{-1}$), as is the case for the sources in this study (see e.g. \citealt{2005MNRAS.362....9Best,2019A&A...622A..17Sabater,2023A&A...676A.102CapettiBrienza}).
Because the radio phase in jetted AGN can be recurrent, quantifying the life cycle of radio galaxies is expected to shed light on our understanding of these topics. Therefore, the crucial first step is to build statistical samples of sources in all stages of the radio life-cycle: young, evolved active, remnant, and restarted.

\textit{Young} radio sources are thought to be represented by peaked spectrum sources (see review by \citealt{2021A&ARv..29....3OdeaSaikia_review} and references therein), with convex-shaped radio spectra. 
Based on the frequency at which the maximum occurs, these sources are classified as high-frequency peakers (HFPs; $\gtrsim$5 GHz), gigahertz peak spectrum sources (GPS; $\sim$1--5 GHz), and compact steep spectrum sources (CSS; $\lesssim$500 MHz). The drop in luminosity below the spectral peak is considered to be caused by the synchrotron self-absorption (SSA, e.g. \citealt{2009AN....330..120Fanti}) or free-free absorption (FFA) due to a dense environment (e.g. \citealt{1999ApJ...521..103Peck,2008A&A...487..885Orienti_Dallacasa,2015ApJ...809..168Callingham, 2015AJ....149...74Tingay}). 
An anticorrelation has been found between the size of the source and the frequency of the peak (see, e.g. \citealt{1990A&A...231..333Fanti}). This relation is most commonly explained as the result of the SSA \citep{1997AJ....113..148O'Dea_Baum,1998PASP..110..493Odea}, with the emission becoming more transparent as the source expands\footnote{The frequency at which the emission peaks is inversely proportional to the source size, either as a result of evolution or by the mechanism for the turnover or both.}.
However, the possibility of explaining the peak spectrum with FFA due to the dense material in which the source is embedded has also been considered \citep{1997ApJ...485..112Bicknell}.
Very likely, both phenomena play a role. These young sources are expected to evolve to the large radio sources in the span of one to hundreds of millions of years \citep{1999A&A...344....7Parma}.

At the other extreme of a radio galaxy's evolutionary track, \textit{remnant} radio sources represent the last stage in the life cycle of radio galaxies. In this stage, the activity stops or dimms significantly, resulting in diffuse and amorphous extended radio emission with no ongoing fuelling.
A young radio source can also be observed in the centre of this remnant radio emission, implying the presence of newly formed compact jets. These newly active radio galaxies are referred to as restarted sources since they have emissions from both the previous and current activity (e.g. \citealt{1974Natur.250..625Willis, 1985A&A...148..243Barthel, 1998PASP..110..493Odea, 2005A&A...443..891Stanghellini, 2010ApJ...715..172Tremblay, 2012A&A...545A..91Shulevski}; \citealt{2019ApJ...875...88Bruni}). 
However, restarted radio sources are mainly known from the studies of `double-double' radio galaxies (DDRG; \citealt{2000MNRAS.315..371Schoenmakers}). These galaxies contain two pairs of distinct lobes on opposite sides of the host galaxy and are, therefore, the easiest to identify based on their morphology alone \citep{2000MNRAS.315..381Kaiser_DDRG,2003ApJ...590..181Saripalli_DDRG,2006MNRAS.366.1391Saikia_DDRG,2009ASPC..407..137Jamrozy_DDRG,2011MNRAS.410..484Brocksopp_DDRG,2012MNRAS.424.1061Konar_DDRG,2019A&A...622A..13Mahatma_restarted}.
A study expanding the selection of candidate restarted radio galaxies beyond DDRGs has been presented by \cite{2012ApJS..199...27Saripalli}. The authors included, as restarted, sources with elongated or unusually bright radio cores compared to their total radio power.
Using this approach, they analysed 119 extended sources from the Australia Telescope Low Brightness Survey \citep{2010MNRAS.402.2792Subrahmanyan_ATLBS}. The study found that 24\% of these sources show signs of restarted radio emission, indicating a relatively short remnant phase (a few percent of their lifetime) based on the comparison with the remnant sample (3\%).
Other studies have used the information on the spectral index distribution within the radio lobes to identify restarted activity. This approach allows the detection of both aged plasma in the extended regions and a new cycle of activity in the central part of the source (\citealt{1982ApJ...257..538Burns, 1994ApJ...421L..23Roettiger, 2020A&A...638A..29Brienza388, 2021A&A...648A...9Morganti_resolvedsi,2021Galax...9...88Morganti_resolved_si2}). 

Although the number of known restarted radio sources has grown in recent years, the studies mentioned above had limitations regarding the diversity of parameters used to identify candidate restarted radio sources and/or the telescope's capabilities to detect e.g. low-surface brightness radio emission. A recent study conducted by \cite{2020A&A...638A..34Jurlin} has overcome some of these limitations by expanding the study to low frequencies, utilising the new possibilities offered by the Low Frequency Array (LOFAR, \citealt{2013A&A...556A...2VanHaarlem}) telescope.
Their goal was to identify restarted candidates in the sample of 156 radio galaxies\footnote{Two sources have been rejected from the original sample of 158 sources. For details, see \citealt{2020A&A...638A..34Jurlin} and \citealt{2021Galax...9..122Jurlin}.} larger than 60 arcsec selected in the Lockman Hole field (LH; \citealt{1986ApJ...302..432Lockman_LH}, covering $\sim$30 deg$^2$). This made it possible not only to compare them to the remnant candidates previously selected by \cite{2017A&A...606A..98B} in the same field, but also to use the remaining sources in the sample as a comparison sample of active radio sources. The availability of these samples allows us to derive fractions of restarted and remnant radio sources and compare them with the models describing the evolution of the remnant radio emission \citep{2017A&A...606A..98B,2017MNRAS.471..891Godfrey,2018MNRAS.475.2768Hardcastle,2020MNRAS.496.1706Shabala}.
Using both morphological and spectral criteria to identify restarted radio sources at different stages, \cite{2020A&A...638A..34Jurlin} classified 13 -- 15\% of the radio galaxies in their sample as restarted candidates.
The criteria used were namely (1) the high Core Prominence\footnote{Core prominence is defined as the ratio between the flux density of the core and the total flux density of the source, $\rm CP= S_{\rm core}\rm \slash{S_{\rm total}}$.} (CP$_{\rm 1.4~GHz}$ > 0.1) combined with low-Surface Brightness (SB$_{\rm 150~MHz}$ $< 50$ mJy arcmin$^{-2}$) of the extended emission, taken as an indication of a possible fading structure resulting from a previous epoch of activity; (2) the Steep Spectrum of the Central region\footnote{We use the terms ``central region'' or ``nuclear region'' throughout this paper to describe an area with an angular size of $\sim$6$^{\prime\prime}$ that coincides with the optical counterpart and any potential unresolved core and/or jets.}(SSC; $\rm \alpha_{150~MHz}^{1.4~GHz}$ $\geq$ 0.7, suggesting the presence of sub-kpc activity instead of only a flat-spectrum core; $\alpha$ is defined as $S_{\nu}\propto\nu^{-\alpha}$); and (3) a visual inspection. The visual inspection allows us to identify sources like DDRGs and, in general, restarted sources which were not selected in the automatic way described above but where the inner jets are resolved (see \citealt{2020A&A...638A..34Jurlin} for more details on the criteria).

Following the criteria outlined in the previous paragraph, the finding of a larger fraction of restarted candidates (between 13 and 15\%), compared to candidate remnant radio sources ($\sim$7\%; \citealt{2017A&A...606A..98B,2021A&A...653A.110Jurlin}), suggested that the restarted phase can occur after a relatively short remnant phase, i.e. resulting in the diffuse low-SB structure being still visible  \citep{2020A&A...638A..34Jurlin,2021A&A...648A...9Morganti_resolvedsi}. The study by \cite{2020A&A...638A..34Jurlin} represents the expansion of the above-mentioned work by \cite{2012ApJS..199...27Saripalli} to lower frequencies (150 MHz). The selection of restarted candidates described above was conducted using the LOFAR image with a resolution of 6.00$^{\prime\prime}$ $\times$ 6.00$^{\prime\prime}$ at 150 MHz (hereafter, LOFAR6 image; \citealt{2021A&A...648A...1Tasse_DEEP_FIELDS}).

The next step for a better characterisation of the restarted radio galaxies includes resolving structures such as jets or lobes in the central regions, in addition to the low-SB lobes at large scales, and searching for a peaked spectrum of the central structures typical of young radio galaxies.
These studies have been done so far only on a limited number of sources (see e.g. \citealt{2010evn..confE..89ParmaVLBI}), and without the availability of a comparison sample.
To address this limitation, in this paper, we take advantage of the new high-resolution (0.38$^{\prime\prime}$ $\times$ 0.30$^{\prime\prime}$) LOFAR image at 150 MHz of the central 6.6 deg$^2$ of the LH, obtained using the international stations (hereafter, LOFAR-IB image; \citealt{2022NatAs.tmp...24Sweijen}). This is the first high-resolution full field image at 150 MHz, allowing the study of nuclear regions at such a low frequency in a systematic way.

The LOFAR-IB image allows us to visually investigate whether the bright central regions of the restarted candidates selected by \cite{2020A&A...638A..34Jurlin} using the CP and SSC criteria contain sub-arcsecond jets.
The new LOFAR-IB image also allows us to derive the central region's flux density without contamination of extended emission from the lobes, which can be quite prominent at low frequencies with lower resolution (e.g. LOFAR6 image) but is not the case at high frequencies, where the extended emission falls below the detection threshold or is resolved out. Therefore, in addition to the LOFAR-IB image we use images at 1.4 and 3 GHz. This allows us to further trace the properties of the radio spectrum of the central region from low to high frequencies.
However, this is possible only for a subset of sources presented in \cite{2020A&A...638A..34Jurlin}, as only the central 6.6 deg$^2$ of the LH has been covered at this high spatial resolution (0.38$^{\prime\prime}$ $\times$ 0.30$^{\prime\prime}$).

The paper is structured as follows.
In Sect.~\ref{sec:sample}, we describe the sample studied in this work and the criteria used to select it. 
The description of the LH radio data is given in Sect.~\ref{sec:Data}. In Sect.~\ref{Sec:analysis}, we describe the properties of the `central region' studied here and the approach in the analysis. The results are presented in Sect.~\ref{Sec:Results}, divided into the central regions' detections and morphologies, their flux densities, and spectral indices. We discuss the results in Sect.~\ref{sec:Discussion}. A summary and conclusions follow in Sect.~\ref{sec:summary_and_conclusions}.
The cosmology adopted throughout the paper assumes a flat universe and the following parameters: $\rm H_{0} = 70$ $\rm km$ $\rm s^{-1}$ $\rm Mpc^{-1}$, $\rm \Omega_{\Lambda} =0.7$, $\rm \Omega_{M} =0.3$.

\section{The sample} \label{sec:sample}
This work studies the nuclear regions of 35 radio galaxies in the LH extragalactic field.
These 35 radio galaxies are located in the central 6.6 deg$^2$ of the LH, the region covered by the LOFAR-IB image, and represent a subset of the 156 extended radio sources with sizes > 60$^{\prime\prime}$ selected by \cite{2020A&A...638A..34Jurlin} in the full extent of the LH region ($\sim$30 deg$^{\rm 2}$). Among the 35 radio sources are six restarted candidates and one remnant radio source, while the remaining 28 sources represent evolved active radio galaxies.

Three of the six restarted candidates were originally selected based on high CP and low-SB criteria (J104113+580755, J104204+573449, J105436+590901). Another restarted candidate was selected using resolved spectral index maps (J104842+585326; \citealt{2021A&A...648A...9Morganti_resolvedsi}). The remaining two candidate restarted radio sources, J104809+573010 and J104912+575014, were originally noted to have ultra-steep spectral properties \citep{2016MNRAS.463.2997Mahony,2021A&A...648A...9Morganti_resolvedsi}. In \cite{2020A&A...638A..34Jurlin}, they were selected based on the morphology (J104809+573010) and high CP and low-SB (J104912+575014). The one remnant radio source in our sample, J104622+581427, was selected based on the ultra-steep resolved spectral index map \citep{2021A&A...648A...9Morganti_resolvedsi}. 

The remaining 28 sources, considered evolved active radio sources, are used to compare their properties in higher resolution images to restarted and remnant candidates. We also inspect whether any restarted candidate is among these 28 radio sources missed by the criteria mentioned here and described in detail in respective papers. 

\section{Radio imaging data} \label{sec:Data}
In this work, we use the LOFAR6 and LOFAR-IB images at 150 MHz, combined with publicly available radio images at frequencies up to 3 GHz. For one restarted candidate (J104113+580755), dedicated observation at 3 GHz were obtained with the \textit{Karl G. Jansky} Very Large Array (VLA) in the A-array configuration. Because this was the only source observed of our programme, we present the results in the Appendix but we include them in the discussion of the paper.
In addition, for another source (J104208+592018) we use archival data at 6 GHz, also obtained with the \textit{Karl G. Jansky} VLA in the A-array configuration and presented in \cite{2021A&A...653A.110Jurlin}. We describe these data in the Appendix.

To display the large-scale morphology of our sources, we also use the LOFAR image at 150 MHz with a resolution of 18.65$^{\prime\prime}$ $\times$ 14.67$^{\prime\prime}$ (hereafter, LOFAR18), which best enhances the extended emission \citep{2016MNRAS.463.2997Mahony}. 

\subsection{High-resolution image at 150 MHz: LOFAR-IB} \label{sec:Data/high-res_ilt}
We make use of the LOFAR-IB image obtained by \cite{2022NatAs.tmp...24Sweijen}. This image exploits the LOFAR international stations together with the Dutch array and reaches a spatial resolution of 0.38$^{\prime\prime}$ $\times$ 0.30$^{\prime\prime}$. It covers the 6.6 $\rm deg^{2}$ of the LH region, centred at $\alpha$ =10h45m00s, $\delta$ =+58d05m00s, and reaches a sensitivity of 25 $\rm \mu$Jy beam$^{-1}$ near the phase centre. The initial calibration was performed using the strategy described in \cite{2022A&A...658A...1Morabito}, followed by further calibration and imaging of the full field, full details of which are in \cite{2022NatAs.tmp...24Sweijen}. 
In total, 2430 sources are detected above 5$\sigma$, and given the high resolution and sensitivity, these can be characterised more precisely than it was previously possible at such low frequency. 

The LOFAR-IB images of the six restarted and one remnant candidate can be seen in Figs.~\ref{fig:restarted} and~\ref{fig:remnant}, respectively. The LOFAR-IB images of the remaining 28 active sources in the sample are shown in Fig.~\ref{fig:COMPARISON}. The inset in all these figures is zoomed in on the expected position of the AGN and shows 3$\sigma_{\rm LOFAR-IB}$ contours in purple.
The value of $\rm \sigma_{LOFAR-IB}$ is measured separately for each source and is, therefore, also referred to as $\rm \sigma_{local}$.

\subsection{Publicly available images at 1.4 and 3 GHz} \label{sec:Data/other}
In addition to the LOFAR6 and LOFAR-IB images at 150 MHz and our observations with the VLA at 3 and 6 GHz, we inspected the publicly available images at frequencies higher than 150 MHz to derive spectral indices of the central regions of the sources.
In particular, we include in the analysis the \textit{Karl G. Jansky} VLA Sky Survey (VLASS; \citealt{2020PASP..132c5001LacyVLASS}), at 3 GHz with a resolution of approximately 2.81$^{\prime\prime}$ $\times$ 2.44$^{\prime\prime}$ (typical resolution of the full survey being 2.5$^{\prime\prime}$ $\times$ 2.5$^{\prime\prime}$) and an reported average single-epoch rms noise of 120 $\rm \mu$Jy beam$^{-1}$. The sources in our sample were observed with the first half of the second epoch of VLASS observations (VLASS 2.1).
Furthermore, we use the Faint Images of the Radio Sky at Twenty-cm (FIRST; \citealt{1995ApJ...450..559BeckerFIRST}) at 1.4 GHz with a resolution of 5.40$^{\prime\prime}$ $\times$ 5.40 $^{\prime\prime}$, and a reported average rms noise of 150 $\rm \mu$Jy beam$^{-1}$.

Table~\ref{tab:surveys} summarises the images used in this work and the median rms noise as measured in this study. The difference in spatial resolution between the surveys and the possible implications are considered when discussing the spectral indices in Sects.~\ref{Sec:Results} and~\ref{sec:Discussion}. 

\begin{table} [h]
\caption{Summary of radio survey properties.}
    \label{tab:surveys}
    \centering
    \begin{tabular}{l|c|c|c} \hline \hline
    Image       & Frequency & Resolution            & Median rms noise     \\
    Name        & [MHz]     & [$^{\prime\prime}$]   & [$\rm \mu$Jy beam$^{-1}$]     \\  \hline
    LOFAR-IB    & 150       & 0.38 $\times$ 0.30    & 50                            \\
    LOFAR6      & 150       & 6.00 $\times$ 6.00    & 100                           \\
    LOFAR18     & 150       & 18.65 $\times$ 14.67  & 210                           \\
    FIRST       & 1400      & 5 .40 $\times$ 5.40   & 150                          \\
    VLA*        & 3000      & 1.08 $\times$ 0.69    & 60                            \\
                & 6000      & 0.29 $\times$ 0.23    & 10                            \\
    VLASS       & 3000      & 2.81 $\times$ 2.44    & 150                           \\ \hline \hline
    \end{tabular}
    \tablefoot{
    \small
    VLA* represents dedicated observations of one restarted candidate at 3 GHz, presented in Appendix~\ref{sec:Data/high-res_vla} and one source from the active comparison sample at 6 GHz presented in \cite{2021A&A...653A.110Jurlin}. Resolutions of the VLASS images varies slightly for the sources in our sample. Here, we list the representative resolution of the sample.}
\end{table}

\section{Analysis}\label{Sec:analysis}
\subsection{Properties and the identification of the central region} \label{Sec:analysis/central_region}
In this study, we focus on the properties of the central regions of our sources.
With `central region', we refer to the region with a size of $\sim$6$^{\prime\prime}$ centred on its AGN. This size corresponds to the resolution of the LOFAR6 image used by \cite{2020A&A...638A..34Jurlin} to select the sample. The location of this central region was determined based on the visual inspection of radio images and the optical identification of the host galaxy in the Sloan Digital Sky Survey (SDSS, \citealt{2017AJ....154...28Blanton_SDSS)}).
The detailed procedure of determining the central region and the optical counterparts is described in \cite{2021A&A...653A.110Jurlin}. This central region can include, in the LOFAR-IB image, a single or multiple components. Each of the detections individually is considered resolved in the LOFAR-IB image, if it satisfies the criteria presented in Sect.~\ref{Sec:analysis/approach}. 

Out of the 35 sources in our sample, 22 ($\sim$63\%) have an SDSS redshift. The vast majority of these redshifts range between 0.3 and 0.8 (see Table.~\ref{tab:optical}) corresponding to a range in linear scales of $\sim$ 27 to 45 kpc in the LOFAR6 and $\sim$ 1.3 to 2.3 kpc in the LOFAR-IB images.
Therefore, the central region is expected to include a combination of the actual core and the beginning of the jet or expanding new lobes in the case of a restarted radio source. 

In those sources in which the position of the central region was less certain, due to no optical identification and/or lack of unresolved emission which could indicate the presence of a core in LOFAR6 radio images, we inspect the new LOFAR-IB images. For three of these sources with uncertain identification (J103913+581445, J104655+572302, and J104819+573515), we could detect an unresolved component in the LOFAR-IB image. We adjusted their positions according to the LOFAR-IB and VLASS detections. New positions are noted with a pink inset in Fig.~\ref{fig:COMPARISON}.

\subsection{Analysis of the central region} \label{Sec:analysis/approach}
For the analysis conducted in this study, we use the flux densities of detections within the central regions or upper limits. First, we outline our approach for analysing the LOFAR-IB image. Then we describe the flux densities obtained from other radio images and, in the last paragraph, we describe the estimate of the flux density uncertainties.

To accurately determine the flux densities of detections in the LOFAR-IB image, we assessed whether each detection within the central region was resolved. This involved fitting a Gaussian in CASA to each LOFAR-IB detection in the central regions of our sources. We classified detections as unresolved if their minor and major axes were comparable (inside the errors) to the beam size. Additionally, we considered a LOFAR-IB detection as unresolved if its integrated flux density and peak flux density were consistent within the uncertainties\footnote{Depending on where the sources are with respect to the phase centre, time and bandwidth smearing will affect them, resulting in both reduced peak intensity and increased size of the detections. Therefore, this work's number of resolved detections represents an upper limit, as smearing may result in artificially resolved sources.}. 

Based on the aforementioned criteria, if a detection (or multiple detections) within the central region is unresolved, we utilise the peak flux density to calculate the spectral index and generate spectral index plots. Conversely, if the detection (or multiple detections) in the LOFAR-IB image is resolved, we measure the flux density by integrating over the area within the LOFAR-IB 3$\sigma_{local}$ contours. If the central region consists of multiple unresolved or resolved detections, their flux densities are summed. Both integrated and peak flux densities were measured using CASA inside 3$\sigma_{LOFAR-IB}$ contours. Peak flux density was determined as the maximum-pixel value at the detection's position.
For the other radio images with lower resolution (LOFAR6, FIRST, and VLASS; see Table~\ref{tab:surveys}), we measure the peak flux densities of the detections within their central regions. While some detections in FIRST and VLASS images are resolved at their respective resolutions, none of those resolved in the LOFAR-IB image are resolved within the central 6$^{\prime\prime}$. Therefore, using their integrated flux densities for the analysis presented in this paper would be misleading.

The total uncertainty ($\Delta$$S$) on the flux densities ($S$) was computed by combining the flux density scale calibration uncertainty ($\Delta$$S_c$) and the image rms noise ($\sigma$) in quadrature, multiplied by the flux density integration area in beam units ($A_{\rm int}$),
\begin{equation}
    \Delta S = \sqrt{(S \cdot \Delta S_{c})^2 + (\sigma \cdot A_{int})^2} .
\end{equation}

For $\Delta$$S_c$ in LOFAR6 and LOFAR-IB images we conservatively assume values of 11\% \citep{2019A&A...622A...1Shimwell} and 20\% \citep{2022NatAs.tmp...24Sweijen}, respectively.
For the two surveys conducted with the VLA at 1.4 and 3 GHz (FIRST and VLASS), we assume a value of 5\% and 10\%, respectively\footnote{For the flux density scale error in the VLASS survey see \href{https://library.nrao.edu/public/memos/vla/vlass/VLASS_013.pdf}{https://library.nrao.edu/public/memos/vla/vlass/VLASS\_013.pdf}.} \citep{1995ApJ...450..559BeckerFIRST,2017ApJS..230....7Perley}. For dedicated VLA observations, we assume an uncertainty of 5\% \citep{2017ApJS..230....7Perley}.

\section{Results}\label{Sec:Results}
In this section, we describe the results regarding the central region of the sources in our sample. We first discuss the number of detections, then the properties of the various groups of objects in our sample, their morphologies in the LOFAR-IB image, their flux densities, and their spectral properties. The results presented in this section are summarised in Table~\ref{tab:flux_si_table}.
\begin{table*}[h!]
\caption{Table with flux densities and spectral indices of the central region.}
\label{tab:flux_si_table}
\centering
\resizebox{\textwidth}{!}{\begin{tabular} {l c c c c c c c c c}
\hline\hline
 Source name                   & Number of         & $\rm S_{peak}$    & $\rm S_{peak}$    & $\rm S_{int}$             & $\rm S_{peak}$    & $\rm S_{peak}$     & $\rm \alpha^{1.4~GHz}_{150~MHz;~LOFAR6}$  & $\rm \alpha^{1.4~GHz}_{150~MHz;~LOFAR-IB}$    & $\rm \alpha^{3~GHz}_{1.4~GHz}$    \\
                               & components        & LOFAR6 [mJy]      & LOFAR-IB [mJy]    & LOFAR-IB [mJy]            & FIRST [mJy]       & VLASS [mJy]        &                               &                                   &                           \\ \hline \hline
 \multicolumn{10}{l}{\textbf{Remnant radio source}}                                                                                                                                                           \\   \hline
 J104622+581427                & 0          & < 1.79        & < 0.11            & -                        & < 0.36            & < 0.43            & -                  & -                 & -      \\   \hline
 \multicolumn{10}{l}{\textbf{Six restarted candidates}}                                                                                                                                                         \\   \hline
 J104113+580755$^{\dagger}$    & 2 (C+J)    & 5 $\pm$ 1     & 1.2 $\pm$ 0.5     & 2.6 $\pm$ 0.6$^{\star}$  & 2.6 $\pm$ 0.2     & 3.0 $\pm$ 0.3     & 0.3 $\pm$ 0.1      & 0.0 $\pm$ 0.1   &-0.2 $\pm$ 0.2     \\
 J104204+573449$^{\dagger}$    & 1 (EL)     & 6 $\pm$ 1     & 0.6 $\pm$ 0.3     & 1.9 $\pm$ 0.5$^{\star}$  & 1.4 $\pm$ 0.2     & 1.1 $\pm$ 0.2     & 0.6 $\pm$ 0.1      & 0.2 $\pm$ 0.1     &  0.3 $\pm$ 0.3    \\
 J104809+573010                & 1 UR       & < 0.91        & 0.11 $\pm$ 0.03$^{\star}$ & 0.05 $\pm$ 0.02  & < 0.42            & < 0.35            & -                  & > -0.58           & -                 \\
 J104842+585326                & 0          & < 0.23        & < 0.11            & -                        & < 0.42            & < 0.50            & -                  & -                 & -                 \\
 J104912+575014$^{\dagger}$    & 1 RC       & 9 $\pm$ 2     & 1.8 $\pm$ 0.5     & 4.2 $\pm$ 0.9$^{\star}$  & 1.6 $\pm$ 0.2     & 1.1 $\pm$ 0.2     & 0.8 $\pm$ 0.1      & 0.4 $\pm$ 0.1   & 0.5 $\pm$ 0.3     \\
 J105436+590901$^{\dagger}$    & 1 (EL)     & 4 $\pm$ 1     & 1.6 $\pm$ 0.5     & 3.3 $\pm$ 0.8$^{\star}$  & 4.8 $\pm$ 0.3     & 3.6 $\pm$ 0.4     & -0.1 $\pm$ 0.1     & -0.2 $\pm$ 0.1  & 0.4 $\pm$ 0.2     \\   \hline   
 \multicolumn{10}{l}{\textbf{Twenty-eight active comparison sources}}                                                                                                                                           \\   \hline
 J103753+571812                & 0          & 1.2+/-0.3     & < 0.18            & -                         & < 0.45            & < 0.47            & > 0.44            & -                 & -                 \\ 
 J103803+581833$^{\dagger}$    & 3 (C+2J)   & 4 $\pm$ 1     & 0.6 $\pm$ 0.1$^{\star}$   & 0.4 $\pm$ 0.1     & 0.7 $\pm$ 0.2     & < 0.47            & 0.8 $\pm$ 0.2     & -0.1 $\pm$ 0.1    & > 0.59            \\
 J103840+573649$^{\dagger}$    & 1  RC      & 2.8 $\pm$ 0.8 & 0.5 $\pm$ 0.3     & 1.4 $\pm$ 0.4$^{\star}$   & 0.6 $\pm$ 0.2     & 0.5 $\pm$ 0.1     & 0.7 $\pm$ 0.2     & 0.3 $\pm$ 0.2     & 0.5 $\pm$ 0.5     \\
 J103856+575247                & 1  UR      & 18 $\pm$ 4    & 0.8 $\pm$ 0.2$^{\star}$    & 1.2 $\pm$ 0.2    & 3.3 $\pm$ 0.4     & 2.2 $\pm$ 0.3     & 0.7 $\pm$ 0.1     & -0.6 $\pm$ 0.1    & 0.5 $\pm$ 0.2     \\
 J103913+581445$^{\dagger}$    & 3 (C+2J)   & < 10.42       & 1.7 $\pm$ 0.7     & 3.0 $\pm$ 0.8$^{\star}$   & < 0.36            & < 0.38            & -                 & > 0.95            & -                 \\
 J103953+583213                & 1  UR      & < 1.86        & 0.6 $\pm$ 0.1$^{\star}$    & 0.7 $\pm$ 0.2    & 1.3 $\pm$ 0.2     & 0.9 $\pm$ 0.2     & < 0.17            & -0.4 $\pm$ 0.1    & 0.5 $\pm$ 0.3     \\      
 J104130+575942$^{\dagger}$    & 2 (2J)     & 33 $\pm$ 9    & 2 $\pm$ 3         & 27 $\pm$ 6$^{\star}$      & 9.6 $\pm$ 0.5     & 3.7 $\pm$ 0.4     & 0.5 $\pm$ 0.1     & 0.5 $\pm$ 0.1   & 1.3 $\pm$ 0.2     \\     
 J104208+592018                & 1 UR       & 3.4 $\pm$ 0.8 & 0.5 $\pm$ 0.1$^{\star}$   & 0.7 $\pm$ 0.2     & 0.5 $\pm$ 0.2     & 0.9 $\pm$ 0.2     & 0.8 $\pm$ 0.2     & 0.0 $\pm$ 0.2     & -0.6 $\pm$ 0.4    \\ 
 J104223+575026$^{\dagger}$    & 2 (2J)     & 1.9 $\pm$ 0.5 & 0.3 $\pm$ 0.2     & 0.9 $\pm$ 0.3$^{\star}$   & < 0.42            & < 0.43            & > 0.68            & > 0.36            & -                 \\      
 J104320+585621$^{\dagger}$    & 1 (EL)     & 4.5 $\pm$ 0.9 & 0.4 $\pm$ 0.2     & 1.0 $\pm$ 0.3$^{\star}$   & 1.0 $\pm$ 0.2     & 0.8 $\pm$ 0.2     & 0.7 $\pm$ 0.1     & 0.0 $\pm$ 0.1     & 0.3 $\pm$ 0.4    \\      
 J104352+581327$^{\dagger}$    & 1 (EL)     & 7 $\pm$ 2     & 0.9 $\pm$ 0.4     & 2.2 $\pm$ 0.5$^{\star}$   & 2.1 $\pm$ 0.2     & 1.4 $\pm$ 0.2     & 0.5 $\pm$ 0.1     & 0.0 $\pm$ 0.1     & 0.5 $\pm$ 0.2     \\      
 J104630+582745$^{\dagger}$    & 2 (2J)     & 13 $\pm$ 4    & 1 $\pm$ 1         & 6 $\pm$ 2$^{\star}$       & 2.3 $\pm$ 0.2     & 1.0 $\pm$ 0.2     & 0.8 $\pm$ 0.1     & 0.4 $\pm$ 0.1     & 1.1 $\pm$ 0.3     \\      
 J104655+572302                & 1  UR      & 0.8 $\pm$ 0.2 & 0.25 $\pm$ 0.07$^{\star}$ & 0.20 $\pm$ 0.06  & 0.6 $\pm$ 0.2      & < 0.38            & 0.2 $\pm$ 0.2     & -0.4 $\pm$ 0.2    & > 0.49            \\      
 J104819+573515$^{\dagger}$    & 1 (EL)     & 9 $\pm$ 2     & 4.2 $\pm$ 0.9     & 8 $\pm$ 2$^{\star}$    & 0.8 $\pm$ 0.2     & < 0.37            & 1.1 $\pm$ 0.1     & 1.0 $\pm$ 0.1     & > 0.97            \\
 J104843+572047$^{\dagger}$    & 1 (EL)     & 5 $\pm$ 1     & 1.5 $\pm$ 0.4     & 3.5 $\pm$ 0.7$^{\star}$   & 1.7 $\pm$ 0.2     & 1.1 $\pm$ 0.2     & 0.5 $\pm$ 0.1    & 0.3 $\pm$ 0.1   & 0.6 $\pm$ 0.3     \\      
 J104917+583627                & 0          & < 28.12       & < 0.3             & -                         & < 0.45            & < 0.61            & -                 & -                 & -                 \\      
 J104918+574530$^{\dagger}$    & 1 (EL)     & 1.8 $\pm$ 0.4 & 0.5 $\pm$ 0.2     & 1.0 $\pm$ 0.2$^{\star}$   & < 0.42           & < 0.45             & >  0.64           & > 0.38            & -                 \\      
 J105117+583832                & 0          & < 1.98        & < 0.12            & -                         & < 0.45            & < 0.43            & -                 & -                 & -                 \\   
 J105132+571115$^{\dagger}$    & 1 (EL)     & 8 $\pm$ 2     & 0.9 $\pm$ 0.5     & 3.3 $\pm$ 0.8$^{\star}$   & 3.0 $\pm$ 0.4     & 2.0 $\pm$ 0.2     & 0.5 $\pm$ 0.1     & 0.0 $\pm$ 0.1     & 0.6 $\pm$ 0.2     \\   
 J105137+592005                & 0          & < 0.77        & < 0.18            & -                         & < 0.45            & < 0.41            & -                 & -                 & -                 \\   
 J105160+570009                & 0          & < 3.46        & < 0.15            & -                         & 0.6 $\pm$ 0.2     & 0.6 $\pm$ 0.2     & < 0.81            & < -0.60           & 0.0 $\pm$ 0.5    \\   
 J105237+573103                & 0          & < 47.34       & < 0.21            & -                         & < 3.09            & < 0.50            & -                 & -                 & -                 \\   
 J105326+570310                & 0          & < 2.98        & < 0.17            & -                         & < 0.45            & < 0.39            & -                 & -                 & -                 \\   
 J105342+565408                & 0          & < 0.76        & < 0.18            & -                         & < 0.39            & < 0.43            & -                 & -                 & -                 \\     
 J105426+573649$^{\dagger}$    & 3 (C+2J)   & < 23.98       & 1.2 $\pm$ 0.8     & 4 $\pm$ 1$^{\star}$   & < 3.21            & 1.1 $\pm$ 0.2     & -                 & > 0.05            & <  1.41           \\      
 J105435+565500                & 0          & < 3.86        & < 0.20            & -                         & < 0.43            & < 0.52            & -                 & -                 & -                 \\       
 J105456+565220                & 0          & 1.2 $\pm$ 0.3 & < 0.19            & -                         & < 0.51            & < 0.43            & >  0.40           & -                 & -                 \\      
 J105622+575334                & 0          & 1.0 $\pm$ 0.2 & < 0.15            & -                         & < 0.37            & < 0.36            & >  0.44           & -                 & -                 \\       \hline
    \end{tabular}}
    \tablefoot{
    \small
    In Col. 1 are the source names in J2000 coordinates, where `$\dagger$' indicates sources in which the central region is resolved; Col. 2 lists the number of components detected in the central region of each source in the LOFAR-IB image, as well as their components, where `C'=core, `J'=jet, `2J'=two jets, `EL'=elongated, `UR'=unresolved, `RC'=resolved compact; Cols. 3, 4, 5, 6, and 7 are core flux densities measured directly from images (see Sect.~\ref{Sec:analysis/approach}) at 150 MHz with 6.00$^{\prime\prime}$ $\times$ 6.00$^{\prime\prime}$} and 0.38$^{\prime\prime}$ $\times$ 0.30$^{\prime\prime}$ resolution, 1.4 GHz with 5.40$^{\prime\prime}$ $\times$ 5.40$^{\prime\prime}$ resolution, and 3 GHz with 2.81$^{\prime\prime}$ $\times$ 2.44$^{\prime\prime}$ resolution, respectively;  Cols. 8, 9, and 10 are the spectral indices of the central region emission calculated between 150 MHz and 1.4 GHz, and 1.4 and 3 GHz.
\end{table*}

\subsection{Detection and morphologies of the central region} \label{Sec:Results/det_morph}
Out of 35 sources in our sample, 22 have one or more detections inside the central region in the LOFAR-IB image (detections are noted with `$\checkmark$' in the inset of Figs~\ref{fig:restarted},~\ref{fig:remnant}, and~\ref{fig:COMPARISON}). We do not detect emission in the central region of the remnant radio source. We comment on the active comparison and restarted samples in the following two subsections.

\subsubsection{Active comparison sample}\label{Sec:Results/det_morph/active}
Out of 28 sources in the active comparison sample, we detect structure in the central region in 17 of them. Thirteen of them (i.e. 76\%) show resolved central emission at the LOFAR-IB image resolution. They are indicated with ‘$\dagger$’ next to their name in Table~\ref{tab:flux_si_table}. In these objects, we observe a variety of structures. Six sources show multiple components in the central regions, representing a (potential) core and one or two jets (see Col. 2 in Table~\ref{tab:flux_si_table}). Among them is one radio source (J103913+581445) that might, in fact, be restarted DDRG based on its morphology. We detect two prominent, elongated structures in three active sources. Of them, J104130+575942 and J104630+582745 show the basis of the two large prominent jets, while J104223+575026 exhibits small-scale lobes, confined in the central region.
We comment on the identification of components and the nature of these sources in Sect.~\ref{sec:Discussion}.
In addition to the six sources with multiple components mentioned above, another six objects have elongated structures detected in the central region, indicating the presence of jets. The remaining source we consider only slightly resolved according to the criteria described in Sect.~\ref{Sec:analysis/approach}. 

In addition to the 13 sources with resolved detections described in the previous paragraph, there are four sources with single, unresolved detections in the central region. 

Out of the 11 active comparison sources with no detection of the central region in the LOFAR-IB image, J104917+583627 and J105237+573103 exhibit FRII-like morphology in the same image. These sources display hotspots in large-scale lobes (those expanding well beyond the central region).

\subsubsection{Candidate restarted sample}\label{Sec:Results/det_morph/restarted}
All but one of the restarted candidates have their central region detected (5/6; $\sim$83\%). The source with the undetected central region is J104842+585326. 

Among the five restarted candidates with a detected central region, we consider four resolved at the LOFAR-IB image resolution (80\% of the five detections, indicated with ‘$\dagger$’ next to their name in Table~\ref{tab:flux_si_table}). The source with the unresolved central region is J104809+573010. In it, only a very faint (3$\rm \sigma_{local}$) detection in the central region, and an extremely faint ($\sim$2.5$\rm \sigma_{local}$) detection of the jet that follows the morphology of the large-scale structure, are seen in the LOFAR-IB image. Since the faint 3$\rm \sigma_{local}$ detection of the nuclear region in the LOFAR-IB image coincides with the optical host galaxy of this source, we consider it a detection.
Among the four restarted candidates with resolved central regions, two cases stand out based on their morphologies: J104113+580755 and J104204+573449. In the former, two well-resolved structures are observed elongated in the same direction as the large-scale diffuse lobes (see Appendix~\ref{sec:Results/J104113} for a discussion of this object).  
In the case of J104204+573449, an elongated structure is detected, possibly implying the presence of sub-kpc jets. The remaining two sources (J104912+575014 and J105436+590901) show only barely extended detections. We consider them resolved because of their difference in peak and integrated flux densities.

\subsection{Flux densities and radio luminosities of the central regions} \label{Sec:Results/flux}
In order to construct the integrated spectral index plots and determine the spectral index values, we measured flux densities of the central region or determined the upper limits. The details of these procedures are described in Sect.~\ref{Sec:analysis}.
The mark `$\star$' next to the value in Cols. 4 and 5 in Table~\ref{tab:flux_si_table} indicates LOFAR-IB measurements employed in the analysis (integrated or peak flux densities, depending whether or not a detection(s) in the central region is resolved).
The peak flux densities from the remaining surveys used in this work can be seen in Cols. 3, 6, and 7 in the same table.

\begin{figure} []
\centering
        \includegraphics[width=\linewidth] {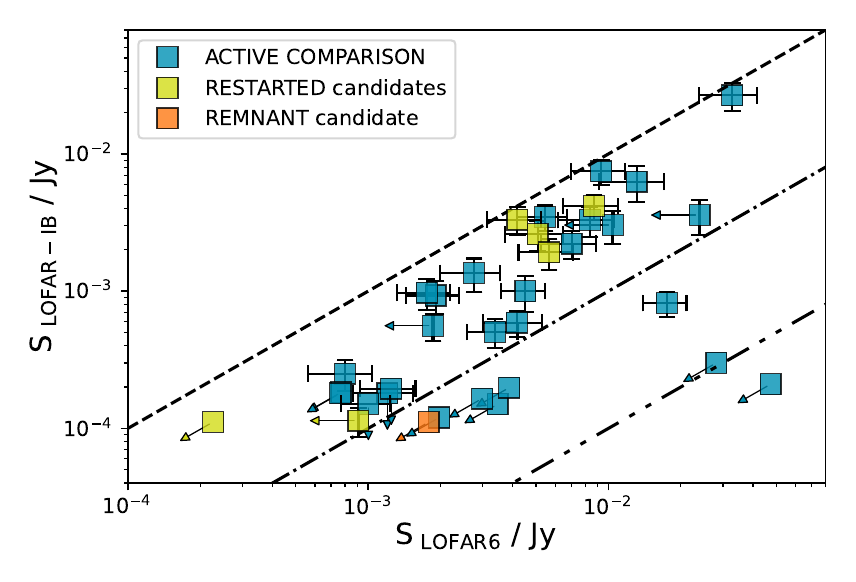}
\caption{LOFAR6 peak flux density vs LOFAR-IB peak or integrated flux density. The only remnant candidate is indicated in orange, restarted candidates in green, and sources from the active comparison sample in blue. Arrows indicate upper limits on flux densities. The dashed line indicates one-to-one relation, the dash-dotted line indicates a factor of 10 difference between the two flux densities at 150 MHz, while the dash-double-dot line indicates a factor of 100 difference. See Sect.~\ref{Sec:Results/flux} for details.}
\label{fig:S_IB_6}
\end{figure}

As a first step, we compared the LOFAR6 and LOFAR-IB flux densities of the central region for each source. This comparison can be seen in Fig.~\ref{fig:S_IB_6}.
This figure shows that the flux densities measured at 150 MHz in the two LOFAR images show, on average, a factor of a few up to $\sim$10 difference (indicated with the dash-dotted line), with a few extreme cases showing a difference of a factor of $\gtrsim$ 100 (J104917+583627 and J105237+573103).
The higher value of the flux density in the LOFAR6 compared to that in the LOFAR-IB image is either due to the extended emission from an active radio source (e.g. faint jets which have not been recovered by the LOFAR-IB images) or the presence of extended diffuse emission, i.e. from lobes.

Using the LOFAR6 flux density results in a systematic tendency for the spectral index of the central regions to be relatively steep. This was already noted in \cite{2020A&A...638A..34Jurlin}, and we will also discuss this in more detail in Sect.~\ref{Sec:Results/si}. Although in some cases, this may reflect the actual presence in the central region of prominent extended structures, e.g. jets and lobes (like we see in J104130+575942 and J104630+582745), we concluded that the high number of steep spectrum cores could suggest a bias introduced by the diffuse emission dominating the spectral index at low frequencies. Further supporting this statement is the fact that the surface brightness sensitivity limit of the LOFAR-IB image is higher (inferior) than that of LOFAR6. In particular, the LOFAR-IB brightness sensitivity limit is estimated to be $\sim$1 mJy arcsec$^{-1}$; see \citealt{2023A&A...671A..85Sweijen}, expected for the remnant or diffuse radio emission. This `contamination' in the LOFAR6 image, but not in LOFAR-IB image, can be seen from the contours in Figs.~\ref{fig:restarted} and~\ref{fig:COMPARISON}, where LOFAR6 detections are embedded in large scale structures, while LOFAR-IB is mostly isolated. This illustrates the importance of having a high spatial resolution image, such as the LOFAR-IB.

Because of the reasons outlined above, in the present study, we consider the LOFAR-IB flux densities (which tend to trace the more compact/small-scale structures) more suitable to be combined with the FIRST and VLASS flux densities for deriving the spectral shape of the central regions.
However, this approach is not without its challenges, as there is a potential risk of missing flux at the high resolution of the LOFAR-IB. We will consider this in the discussion below. 

As one would expect, the number of undetected central regions in the LOFAR-IB image increases as the total flux density in the LOFAR6 image decreases, particularly below $\sim$7 mJy. However, some faint central regions are still detected in the LOFAR-IB images (see Fig.~\ref{fig:S_IB_6}).

\begin{figure} []
\centering
        \includegraphics[width=\linewidth] {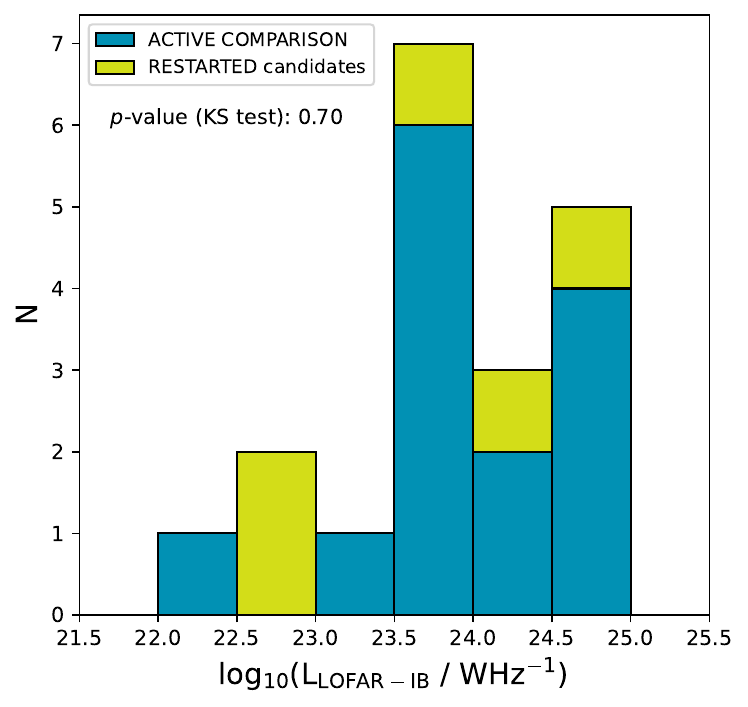}
\caption{The stacked bar chart showing LOFAR-IB radio luminosities of detected central regions at 150 MHz of candidate restarted (green) and comparison active (blue) radio galaxies with SDSS optical counterparts. At the top of the histogram, we report the $p$-value from the Kolmogorov–Smirnov (KS) comparison of two datasets (see, Sect.~\ref{sec:Discussion})}
\label{fig:hist_lum}
\end{figure}

For those sources with detections in their central regions and assigned SDSS optical counterparts, we computed the radio luminosities of their central regions. In Fig.~\ref{fig:hist_lum}, we show the histogram of the distributions of radio luminosities. The sources in our sample have a median radio luminosity of $\rm log_{10}(L_{LOFAR-IB}~/~WHz^{-1})=$ 23.98.

\subsection{Spectral indices of the central regions} \label{Sec:Results/si}
\begin{figure*} [!h]
\centering
    \includegraphics[width=\linewidth] {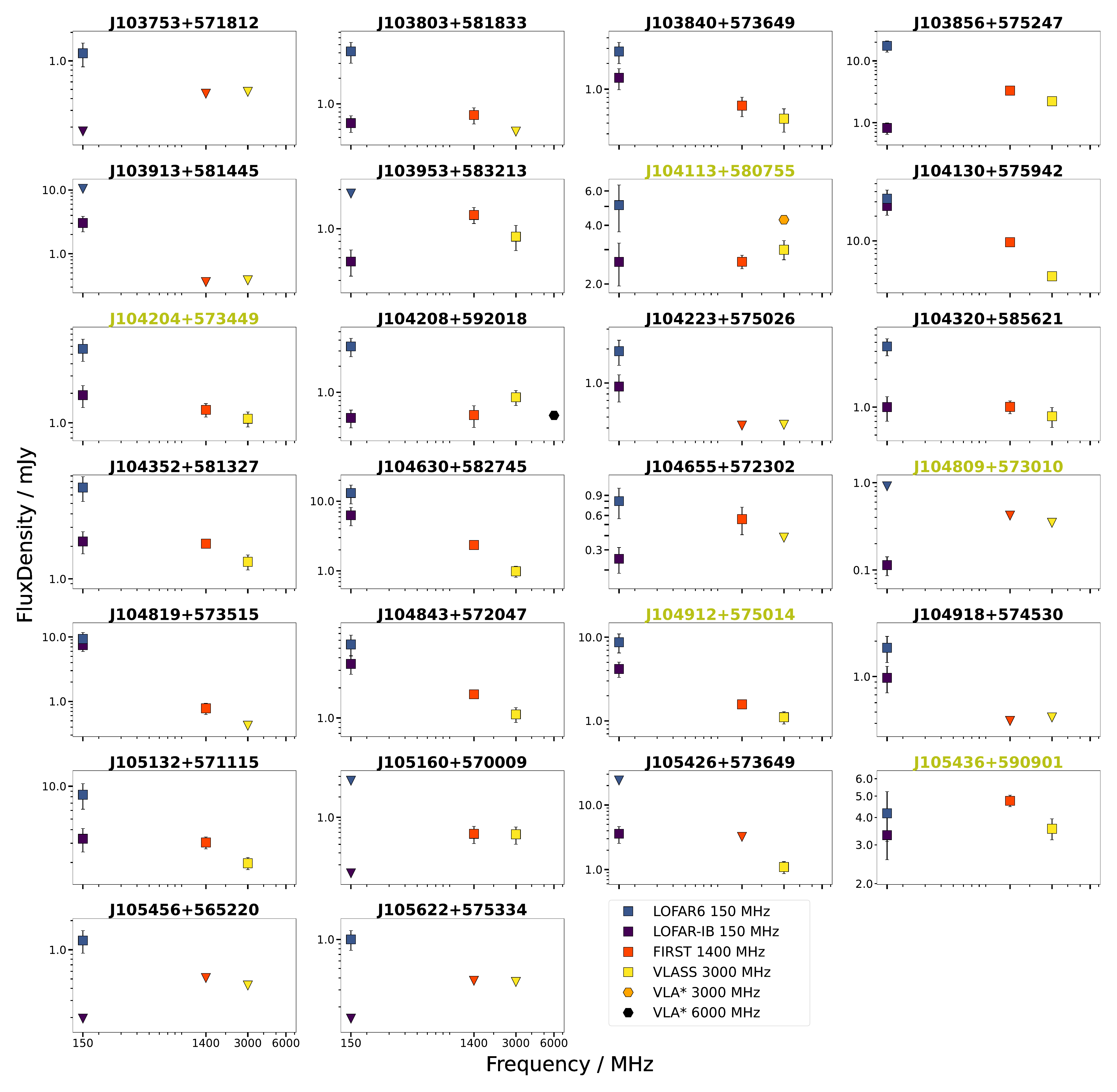}
\caption{Radio spectra of the cores of the 26 sources in the sample. Flux density measurements from LOFAR image at 150 MHz (LOFAR-IB in purple and LOFAR6 in blue), FIRST images at 1.4 GHz (in red), VLASS images at 3 GHz (in yellow), and VLA at 3 GHz and 6 GHz (in orange and black, respectively; only for two sources; see Appendix~\ref{sec:Data/high-res_vla}). Triangles indicate upper limits. Restarted candidates are indicated with a green title, and the rest belong to the active comparison sample.}
\label{fig:spectra}
\end{figure*}

Using the flux density measurements obtained as described in Sect.~\ref{Sec:analysis/approach}, we produced the radio spectra shown in Fig.~\ref{fig:spectra}. These plots show the two measurements at 150 MHz from LOFAR6 and LOFAR-IB, FIRST at 1.4 GHz, VLASS at 3 GHz, and a measurement of the flux density for one restarted candidate at 3 GHz and for one source from the active comparison sample at 6 GHz.
Using these values, we have derived the spectral indices between 150 MHz and 1.4 GHz ($\rm \alpha_{150~MHz}^{1.4~GHz}$), and between 1.4 and 3 GHz ($\rm \alpha_{1.4~GHz}^{3~GHz}$). The derived spectral indices are listed in Table~\ref{tab:flux_si_table} (columns 8, 9, and 10). From Fig.~\ref{fig:spectra} and the spectral index values, we can look for sources that show inverted or peaked spectra, which, as discussed in the introduction, can be used to infer the evolutionary stage of the source.
However, as shown in Fig.~\ref{fig:spectra}, only 26 of the 35 sources ($\sim$74\%) have a detection of the central region (indicated by a square symbol) in at least one of the three frequencies at which our sources are observed\footnote{Throughout this paper, we focus on observations up to 3 GHz while we discuss the observation at 6 GHz only when discussing the source observed at this frequency.}. As a result, we can only calculate spectral indices or spectral index limits for these 26 sources.

As discussed above (Sect.~\ref{Sec:Results/flux}), care should be taken in interpreting the spectral indices because of the difference in spatial resolution between the LOFAR-IB and the high-frequency data (FIRST and VLASS).
Nevertheless, the high resolution and sensitivity of the low-frequency images allow us to trace the small-scale structures in the central regions, which are also the ones more likely to be recovered by the FIRST and VLASS. 
Keeping the spectral index limitations in mind, in the analysis below, we will remark on cases where follow-up observations will be required to confirm the results. 

In addition to the difference in resolution, given that the observations were carried out at different epochs, the intrinsic variability can also be a source of uncertainty in the spectral indices presented here (see, e.g. \citealt{2020ApJ...905...74NylandVARIABILITY,2021MNRAS.501.6139Ross_spectral_variability}). 

Interestingly, despite these uncertainties, the majority ($\sim$82\%) of the sources tend to have a relatively flat-spectrum spectral index in the range [-0.4, 0.5] when looking at the spectral index between 150 MHz (LOFAR-IB) and 1.4 GHz. The flat-spectrum emission is consistent with the typical spectral indices of cores in radio galaxies (see, e.g. \citealt{1979ApJ...232...34Blandford,1984A&A...139...55Feretti, 2008MNRAS.390..595Mullin}).\\

\begin{figure} [!h]
    \includegraphics[width=\linewidth] {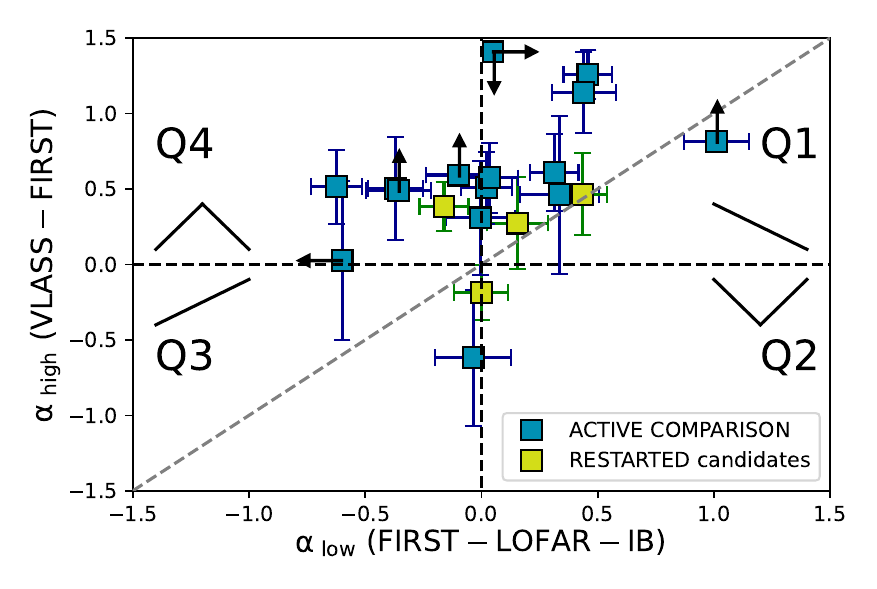}
    \includegraphics[width=\linewidth] {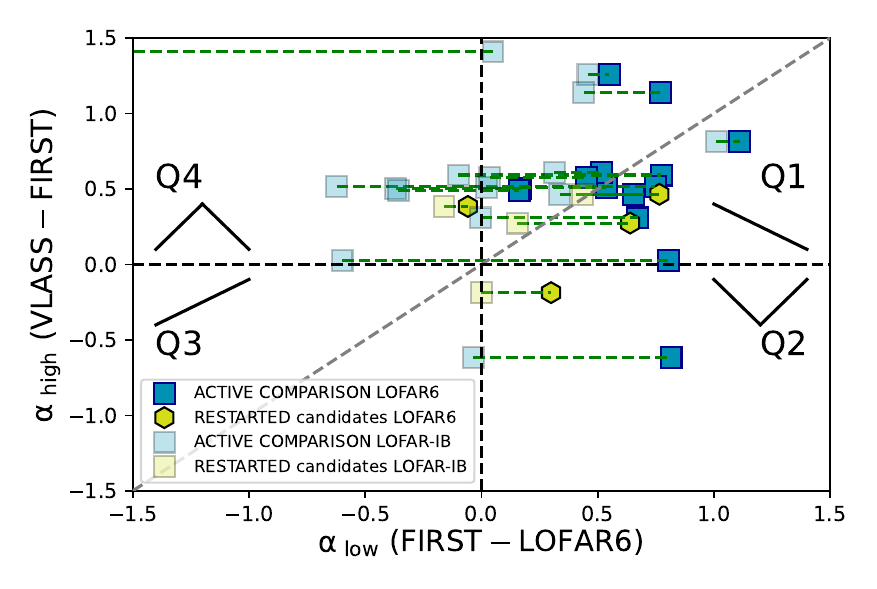}
\caption{$\rm \alpha_{~LOFAR-IB}^{~FIRST}$ vs $\rm \alpha_{~FIRST}^{~VLASS}$. The one-to-one relation of $\rm \alpha_{low}$ and $\rm \alpha_{high}$ is shown with a dashed grey line. The dashed black lines represent spectral indices of zero, dividing the plot in four quadrants (Q1, Q2, Q3, and Q4). The black lines in the corner of each quadrant illustrate the shape of the spectrum. We remind the reader that $\alpha$ is defined as $S_{\nu} \propto \nu^{-\alpha}$.
Arrows indicate upper or lower limits. In the upper panel, we use LOFAR-IB flux density measurements, while in the lower panel, we use LOFAR6 flux densities and compare them to their position in the upper panel.}
\label{fig:alpha_low6_ilt_high}
\end{figure}

However, in this study, we can expand this analysis. Since we have flux densities measured at three distinct frequencies, we can construct a colour--colour plot (see Fig.~\ref{fig:alpha_low6_ilt_high}). In this plot, we can identify the shape of the spectrum from the location of the source in it. 
Seven of the 26 sources have a detection at only one of the three distinct frequencies used in this plot (e.g. either at 150 MHz or 1.4 GHz or 3 GHz; see Fig.~\ref{fig:spectra}). Therefore, these objects cannot be included in the colour--colour plot, and only a limit for the spectral index can be derived. 
Among them is J103913+581445, a possible DDRG, as revealed from the LOFAR-IB morphology of the central region, already mentioned in Sect.~\ref{Sec:Results/det_morph} and further discussed in Sect.~\ref{sec:Discussion}. 

The colour--colour plot obtained using the LOFAR-IB flux densities of the central component is shown in Fig.~\ref{fig:alpha_low6_ilt_high}. Colour--colour diagrams have been used extensively by, e.g. \cite{1977AJ.....82..541Kesteven}, \cite{1986AJ.....91.1011Rudnick}, and more recently, \cite{2022AJ....164..122McCaffrey} and \cite{2022ApJ...934...26Patil_cc_plot} to examine the spectra of samples of radio sources, radio cores, quasars of different radio-loudness, and heavily obscured luminous quasars, respectively. 
This plot has also been used to assess a spectral peak by \cite{2006MNRAS.371..898Sadler,2011MNRAS.412..318Massardi,2016MNRAS.463.2997Mahony,2017ApJ...836..174Callingham}, among others. In it, radio colour--colour space is defined by the spectral index derived between two high frequencies (in our case, 1.4 and 3 GHz; $\rm \alpha_{high}$) and the spectral index derived between two lower frequencies (in our case, 150 MHz and 1.4 GHz; $\rm \alpha_{low}$). 
Therefore, in Fig.~\ref{fig:alpha_low6_ilt_high}, we mark the division in four quadrants indicated by the two perpendicular dashed lines (Q1, Q2, Q3, and Q4).\\ 

For comparison we present in Fig.~\ref{fig:alpha_low6_ilt_high} the same plot obtained using the LOFAR6 flux densities of the central region. It is clear from this plot, consistent with what is shown in the spectral plots of Fig.~\ref{fig:spectra}, that most of the sources move towards steeper values of the spectral index of the central regions. The distribution appears much steeper than what typically expected for the central regions of radio galaxies, suggesting a strong contribution from the diffuse emission which becomes more prominent at low frequencies. We discuss below some specific cases supporting the choice of using the LOFAR-IB flux densities.

Looking at the upper panel in Fig.~\ref{fig:alpha_low6_ilt_high}, the sources in our sample are spread throughout the plot, with no clear separation between the groups. However, the whole sample mainly occupies Q1 and Q4.
In Q1, the majority of the sources have a relatively flat spectral index\footnote{Unlike in \cite{2020A&A...638A..34Jurlin}, where the central region is considered steep if $\rm \alpha$ $>$ 0.7, here we consider the central region steep if $\rm \alpha$ $>$ 0.5 due to it not being contaminated by the extended diffuse emission at 150 MHz, which was the case in \cite{2020A&A...638A..34Jurlin}.} ($\leq$ 0.5) throughout the range of frequencies we have covered. This flat-spectrum central region is consistent with the typical spectral indices of cores in radio galaxies as a result of self-absorption (see, e.g. \citealt{1979ApJ...232...34Blandford,1984A&A...139...55Feretti, 2008MNRAS.390..595Mullin}). The two restarted candidates in Q1 are J104204+573449 and J104912+575014. The two sources (J104130+575247 and J104630+582745) with a particularly steep spectrum are part of the active comparison sample. In those cases, the central region includes prominent jets connected to the ones seen on a large scale. Therefore, in these two cases, the steep spectrum represents the spectrum of the large-scale jets, not that of the newly restarted activity.

Q4 is particularly interesting because it includes objects with a peaked spectrum, meaning they are inverted at low frequencies and steep at high frequencies. For our frequency coverage, the peak of these spectra is likely located at $\sim$1 GHz. We describe two sources in this quadrant in more detail due to their peaked spectra. The first case is the restarted radio source J105436+590901. Intriguingly, unlike the rest of the sources in this quadrant, this source would remain peaked even if the LOFAR6 flux density is used, suggesting that the emission from the entire central region is characterised by a peaked spectrum, with the peak at low frequencies ($\lesssim$ 1 GHz). Recently, there have been studies identifying sources with observed peak frequencies below 1 GHz, referred to as MHz-peaked spectrum (MPS) sources (see, e.g. \citealt{2015MNRAS.450.1477CoppejansMPS1,2016MNRAS.459.2455CoppejansMPS2,2017ApJ...836..174Callingham}). Based on its spectral shape, source J105436+590901 is consistent with being classified as an MPS source. The second case is the source, J103856+575247, which is part of the active comparison sample and shows a peaked spectrum. However, in the case of this source and others like J104655+572302, follow-up observations, with improved resolution, sensitivity, and conducted simultaneously to mitigate potential variability issues, will be necessary to confirm that high-resolution LOFAR-IB imaging has not missed any flux compared to FIRST and VLASS.

The quadrant Q3 corresponds to synchrotron self-absorbed spectra and in this quadrant we would expect to see GPS or HFP sources that are peaking above 3 GHz.
Sources that display convex spectra are located in the second quadrant of Fig.~\ref{fig:alpha_low6_ilt_high}. The convex spectrum category is likely composed of sources that have had multiple epochs of AGN activity, with the peaked-spectrum component above 1 GHz representing recent activity in the core. At the same time, the upturn at frequencies below the turnover is suggestive of diffuse, older emission \citep{1990A&A...232...19Baum,2004A&A...424...91Edwards&Tingay,2007A&A...469..451Torniainen,2010MNRAS.408.1187Hancock}. 
In these types of sources, the peaked component of the spectrum should be interpreted as the radio source being re-started on short timescales, although variability cannot be fully excluded. One source from the active comparison sample (J104208+592018) and one restarted candidate (J104113+580755) are located at the borderline of the second and third quadrants. 

Interestingly, the flux density of J104113+580755 at 3~GHz with resolution roughly matching the LOFAR-IB (see, Appendix~\ref{sec:Data/high-res_vla}) remains the same as from VLASS, therefore confirming the inverted spectrum between low and high frequencies. Also interesting is the fact that the 6~GHz observations of the source J104208+592018, matching the resolution of LOFAR-IB, are consistent with the relatively flat spectrum found from the LOFAR-IB--FIRST and VLASS. Whether the spectrum presents a peak around the 3~GHz, will have to be investigated using multi-frequency observations with matching resolution. These two objects seem to support the choice of using the LOFAR-IB for deriving the spectral properties of the compact structures in the central regions.

\section{Discussion}
\label{sec:Discussion}
This study gives us a first view of the central regions at kpc scales of a sample of radio galaxies at 0.3 arcsec resolution using LOFAR-IB at 150 MHz.
We have used the high-resolution morphology and the spectral index information to explore the evolutionary stage of the 35 sources in the sample initially selected by \cite{2020A&A...638A..34Jurlin}, and classified as remnant, restarted, or active by \cite{2020A&A...638A..34Jurlin} and \cite{2021A&A...648A...9Morganti_resolvedsi} (see Sect.~\ref{sec:sample}).
Below we discuss the morphological and spectral properties of the whole sample.\\

We detect at least one component in the central region of 22 of the 35 radio sources in our sample. Of these 22 sources, 15 have only one component, four have two components, and the remaining three sources have three components. Studies with a similar goal of investigating cores and nuclear regions have been done in the past on the B2 and 3C samples (e.g. \citealt{1987A&A...181..244Parma,1988A&A...199...73Giovannini,1990A&A...227..351DeRuiter,2008MNRAS.390..595Mullin}). However, these studies were conducted on samples at lower redshifts (z $<$ 0.2) than studied here, spanning a broader range in radio luminosities (e.g. 10$^{23}$--10$^{28}$ W Hz$^{-1}$ at 408 MHz in \citealt{1988A&A...199...73Giovannini}, compared to 10$^{24}$--10$^{26}$ W Hz$^{-1}$ at 150 MHz in this work\footnote{For the properties of the whole LH sample, see \cite{2020A&A...638A..34Jurlin}.}). Furthermore, the presence of cores was investigated using higher frequencies than in this work, specifically 5 GHz \citep{1988A&A...199...73Giovannini,1990A&A...227..351DeRuiter}. Despite these differences and considering there is a higher chance of detecting a radio core at a higher frequency, it is interesting to see that the fraction of detections of nuclear regions reported in this work ($\sim$63\%) is comparable to the detections in the previous studies mentioned in this paragraph (between $\sim$65\% and $\sim$80\%). Furthermore, the study investigating the properties of radio galaxies from the B2 sample reports that $\sim$45\% of the sources show unambiguous detection of radio jets at 1.4 GHz \citep{1987A&A...181..244Parma}, which is comparable to the sources in our sample with jets or elongated detections ($\sim$43\%; see Table~\ref{tab:flux_si_table}).\\ 

The morphological inspection of a nuclear region can tell us about the nature of our sources. For example, only in two sources, we detect the beginnings of the large-scale jets (J104130+575942 and J104630+582745). These detections are noticeably elongated and connected to the collimated jets detected in lower-resolution images, confirming the ongoing activity in these sources. The fact that we see the beginnings of large-scale jets in only two sources is surprising because most sources have a radio luminosity typical of Fanaroff-Riley type I radio sources (FRI; \citealt{1974MNRAS.167P..31Fanaroff}). Although the distinction between FRI and FRII (Fanaroff-Riley type II) sources based on their radio luminosities has been questioned (see, e.g. \citealt{2019MNRAS.488.2701Mingo}), the vast majority of the sources in our sample have large-scale morphologies that support their classification as FRIs (see Figs.~\ref{fig:restarted},~\ref{fig:remnant},~\ref{fig:COMPARISON}).

The other sources in our sample with multiple components have either only small-scale jets confined in the central region or core and (unresolved) components.
Thus, either the morphologies of these detections are affected by jet-ISM interaction \citep{1994AJ....108..766Bridle} keeping the jet (temporarily) confined to the inner few kpc region or short-time variability in the fuelling (and consequent emission) originates in discrete blobs of emission. However, it also might be due to the limit in detecting the low-SB emission in the LOFAR-IB image. The latter is likely the case for, e.g., J103803+581833 (see Fig.~\ref{fig:COMPARISON}). In this source, the detections on the opposite side of the assumed core are elongated in the same direction as the large-scale FRI jets/lobes. They are, therefore, more likely to represent higher flux density regions of an active jet.

Interestingly, we discovered some potentially restarted galaxies based solely on morphology. This is the case for J104113+580755, J104223+575026, and J103913+581445 where multiple detections in their central regions do not appear to be the inner part of large-scale structures, since the extended emission is amorphous and/or does not show visible jets. Source J103913+581445 has three detections in its central regions, one a potential core and two lobes symmetric on both sides of the core, reminiscent of the inner structure of a typical DDRG.
Interestingly, two of the three sources mentioned above are in the active sample (J104223+575026 and J103913+581445), meaning that they were missed by the criteria used in \cite{2020A&A...638A..34Jurlin}, emphasising the importance of high-resolution and high-sensitivity images when selecting potential restarted radio sources.

Finally, there are 15 sources in our sample with one detection in the central $\sim$6$^{\prime\prime}$. The majority of these detections are compact. Compact detections for the median redshift of the sources in our sample (z = 0.5, based on the SDSS information) at this resolution (0.38$^{\prime\prime}$ $\times$ 0.30$^{\prime\prime}$) imply a component with a projected linear size of 1.8 kpc. 
On the other hand, those detections showing elongated structure might indicate newly-formed jet(s) and, therefore, a restarted radio activity (e.g. J104204+573449). However, the nature of this elongated component would need to be confirmed by the spectral analysis or images at even higher resolutions. 

For the analysis of the spectral index, focused on assessing a spectral peak, we have constructed the colour–colour plot shown in Fig.~\ref{fig:alpha_low6_ilt_high}.
Two limitations of the data that we use should be kept in mind.
First, the data are not taken simultaneously, and some variability could be expected (e.g. \citealt{2020ApJ...905...74NylandVARIABILITY,2021MNRAS.501.6139Ross_spectral_variability}).
Second, the spatial resolution of the LOFAR-IB images is higher than that of the FIRST and VLASS images. This may appear to be a limitation, but in fact, it gives us a more comparable view of the central region between the low and high frequencies. This is because the LOFAR6 image can be affected by diffuse emission in some cases (see Table~\ref{tab:surveys} and Fig.~\ref{fig:S_IB_6}), which would significantly impact the resulting spectrum of the more compact regions by making it unrealistically steep. As a result, while the resolution mismatch is not ideal, we believe it provides constraints on the spectral properties of the core regions.

In our sample, we detect various spectral shapes of nuclear regions (see Figs.~\ref{fig:spectra} and~\ref{fig:alpha_low6_ilt_high}). They include candidate peaked spectra (e.g. J105436+590901), possibly convex spectra (e.g. J104113+580755), as well as steep spectra (e.g. J104130+575942), all indicating on-going activity (a young source in the first two cases, and jets in the latter).
There are seven sources in which the central region is detected at only one frequency (see Fig.~\ref{fig:spectra}). Therefore, these seven sources are not included in Fig.~\ref{fig:alpha_low6_ilt_high}.
Among them is one interesting source, J103913+581445, which shows a morphology resembling a DDRG with a steep spectral index limit ($>$ 1.03) between 150 MHz and 1.4 GHz. 

Overall, both active and restarted candidates show a similar fraction of resolved central regions (76\% and 80\%, respectively), and there is no statistical difference\footnote{We use the Kolmogorov–Smirnov (KS) comparison of two datasets (a two-sided KS test; \citealt{1933Kolmogorov}) to quantify the statistical difference between the distributions, at the 95\% confidence level. The ranges of the radio luminosity distribution for the evolved active radio sources and restarted candidates are statistically similar with a $p$-value equal to 0.70. Therefore, we cannot reject the null hypothesis that these two samples come from the same distribution.} between the radio luminosities of their central regions (see Fig.~\ref{fig:hist_lum}). The sources in the active comparison sample tend to show more structure in the LOFAR-IB image beyond the central region (e.g. hotspots or jet structures). We find that the central regions of the active comparison sample can show the beginnings of large-scale jets. 
In contrast, the detections in the central regions of restarted candidates are more compact and do not seem connected to the large-scale structure, which is typically more amorphous. We comment on this for some cases from the samples of restarted candidates and active comparison sources in the following two sub-sections.

\subsection{Restarted candidates} \label{sec:discussion/restarted}
The LOFAR-IB observations have confirmed that five of the six restarted candidates have an active nucleus despite the low surface brightness and amorphous emission detected on large scales. 

For one object, namely J104842+585326, we do not detect the central region in the LOFAR-IB image. However, based on the possible optical counterpart reported by \cite{2021A&A...648A...3Kondapally}, \cite{2021A&A...648A...9Morganti_resolvedsi} already propose that this source might be a high redshift (z = 2.1685) FRII source in the process of switching off instead of a restarted source. This hypothesis would be consistent with the non-detection of its central region.
Therefore, we reject this source as a potentially restarted radio source.

Three of the remaining five restarted candidates were originally selected based on the high-CP and low-SB criteria. The initial reason to adopt the CP criterion was based on the expectation that the relatively bright central region (compared to the extended emission) would reveal an extended structure, a multi-component structure, or a possibly peaked radio source when using data of adequate resolutions and sensitivities to investigate these possibilities. The analysis presented here confirms our expectation; see, e.g. J104204+573449 with extended structure, J104113+580755 with multi-component structure, and J105436+590901 with a peak in its radio spectrum around 1 GHz. 
Furthermore, the low-SB extended emission did not reveal any compact component in the high-resolution images, both at high- (VLASS and VLA) and low-frequency (LOFAR-IB), as expected if this emission is from remnant radio plasma surrounding the restarted radio source. 

For the restarted candidate with multiple components in the central region (J104113+580755), the spectral analysis revealed that one detection exhibits a flat spectral index, while another displays a steep one (see Appendix~\ref{sec:Results/J104113}). Furthermore, the flat spectrum component coincides with the optical counterpart. 
Therefore, we are confident that these detections represent a core (flat-spectrum component coinciding with the optical host galaxy) and a jet (steep-spectrum component). 
The jet structure appears limited to the central region, while the large-scale emission is amorphous. We interpret this as resulting from a new activity cycle. 
Even if the asymmetric central structure is affected by some beaming, the presence of symmetric and amorphous low-surface brightness extended structure would still support the presence of a restarted activity giving rise to the central emission.

The remaining two restarted candidates were both originally selected for their large-scale spectral properties (J104809+573010 and J104912+575014).
They were noted to have extreme spectral properties ($\rm \alpha_{150~MHz}^{1.4~GHz}$ $>$ 1.2) by \cite{2016MNRAS.463.2997Mahony} and their restarted nature was further confirmed by the resolved spectral index analysis presented in \cite{2021A&A...648A...9Morganti_resolvedsi}. Additionally, the source J104912+575014 was selected as restarted candidate in \cite{2020A&A...638A..34Jurlin} based on the high CP and low-SB criterion. Its central region is barely resolved in the LOFAR-IB image and described with a steep spectral index, possibly indicating structure on sub-kpc scales. Therefore, images at an even higher resolution than what was available at this work would be needed to study the nuclear region of this source in more detail.
The central regions of these two sources in the LOFAR-IB image do not show extended or multiple components at the resolution and sensitivity of the LOFAR-IB image, indicating new activity. This confirms that criteria based on both morphology and radio spectra are needed to select the full statistical sample of restarted radio sources.

In conclusion, these findings highlight the importance of employing a diverse range of criteria to identify restarted radio sources. The criteria used in the current study, as well as in \cite{2020A&A...638A..34Jurlin}, have proven valuable in identifying additional candidates belonging to this important group of radio sources.

\subsection{Comparison active sample}\label{sec:discussion/active}
Radio galaxies from the active sample show a variety of morphological and spectral properties. Interestingly, three sources in the comparison sample could be identified as possible restarted candidates, thanks to their morphology revealed in the high-resolution image and the derived spectral indices.
Two sources (J103913+581445 and J104223+575026) were identified as candidate DDRG by the morphology identified in the LOFAR-IB image in combination with what is seen at larger scales. The first one, J103913+581445, is also characterised by a steep spectral index in the central region. 
The third is source J103856+575247, where a possible peaked spectrum source is detected in its central region. Higher sensitivity and high-resolution, multi-frequency data will be needed to confirm their nature. 

\subsection{Occurrence and implications for the life-cycle}
Although there is no clear separation in the spectral properties of the groups, the morphological analysis and the spectral index analysis allowed us to characterise the central regions of the sources and helped us find new restarted candidates.

While we reject one restarted candidate (J104842+585326), we affirm the remaining five restarted candidates, supporting the selection presented in \cite{2020A&A...638A..34Jurlin}. In addition, we find three more candidates based on their sub-arcsec morphology and/or spectrum characteristics. As a result of the analysis presented in this paper, the fraction of restarted candidates in the entire LH region is up to 15\%, corroborating the number of restarted candidates found in \cite{2020A&A...638A..34Jurlin}. Confirming the occurrence of restarted candidates is the primary result of the paper because it has implication for the life-cycle of radio AGN.
In particular, our study supports the duration of the inactive phase of our restarting candidates to last a few tens of millions of years \citep{2020A&A...638A..34Jurlin}. This duration is estimated based on the fraction of restarted (up to 15\%) and remnant candidates (9\%; \citealt{2021A&A...653A.110Jurlin,2021A&A...648A...9Morganti_resolvedsi}) in the entire LH region, and modelling presented in \cite{2017A&A...606A..98B} and \cite{2017MNRAS.471..891Godfrey}. For more details, we refer to \cite{2020A&A...638A..34Jurlin}.
The fraction of restarted candidates derived by the observations has been also used for theoretical modelling to predict the evolution of radio sources \citep{2020MNRAS.496.1706Shabala}. In \cite{2020MNRAS.496.1706Shabala}, it is stated that in order to achieve a fraction of restarted radio sources greater than 10\%, a ``power law age'' model characterised by a substantial population of short-lived radio sources is necessary.

The three potential restarted candidates added from the active comparison sample show us that even though the selection based on the high-CP and low-SB has proven to be good in selecting restarted candidates, it has limitation in selecting certain cases of restarted radio sources. However, other restarted candidates can be selected based on their sub-arcsec morphology and/or peaked spectrum, further supporting the criteria presented by \cite{2020A&A...638A..34Jurlin}. High-resolution LOFAR-IB image played a crucial role in this selection.

\section{Summary and conclusions} \label{sec:summary_and_conclusions}
Thanks to LOFAR, in recent years, we have been able to systematically select samples of radio sources in various phases of their life cycles. Furthermore, LOFAR high-resolution imaging at a low frequency with an excellent sensitivity, used in this work, opens a new window into detailed studies of low-luminosity radio sources and their nuclear regions.

In this study, we used information from 150 MHz to 3 GHz to examine the central regions of 35 radio sources in the Lockman Hole field.
Here we list the main conclusions from this work:
\begin{itemize}
    \item We detect radio emission in the central $\sim$6$^{\prime\prime}$ of 22 of the 35 sources (63\%) in our sample (see Sect.~\ref{Sec:Results/det_morph}).
    \item The central region is detected mainly in restarted candidates (83\%), and 80\% of those central regions are, at least slightly, resolved (see Sect.~\ref{Sec:Results/det_morph}).
    \item Among the restarted candidates, there is one source with a peaked radio spectrum (J105436+590901) and one with a convex radio spectrum (J104113+580755) between 150 MHz and 3 GHz (see Sect.~\ref{Sec:Results/si}). These two restarted candidates possibly represent the youngest restarted radio sources.
    \item Sources from the active comparison sample show a larger variety of structures in the LOFAR-IB image. Some of their central regions show the inner part of the large-scale jets (see Sect.~\ref{Sec:Results/det_morph} and Fig.~\ref{fig:COMPARISON}).
    \item Overall, the sources in the comparison active sample show much more structure (often showing hotpots or knots in jets) in the full extent of the radio emission compared to restarted and remnant candidates (see Figs.~\ref{fig:remnant},~\ref{fig:restarted},~\ref{fig:COMPARISON}).
    \item Interestingly, the three objects in the active comparison sample could have a signature of restarted activity (either based on their morphology or their spectral shape): J103856+575247, J103913+581445, and J104223+575026 (see Sect.~\ref{sec:Discussion}). 
    \item We confirm the prevalence of restarted candidates over remnants and the duration of the inactive phase of our restarting candidates to last a few tens of millions of years as reported in \cite{2020A&A...638A..34Jurlin} (see Sect.~\ref{sec:Discussion}).
    \item The combination of the high-CP and low-SB has proven to be a good criterion in selecting restarted radio sources (see Sect.~\ref{sec:Discussion}).
    \item In order to select restarted candidates in various evolutionary stages, criteria based on both morphological and spectral properties need to be employed (see Sect.~\ref{sec:Discussion}).
    \item High-resolution images are a crucial part of selecting restarted radio sources and bring us closer to understanding the radio life-cycle.
\end{itemize}

Overall, the present study builds upon powerful tools opening promising perspectives for exploring the sources' evolutionary status in more detail. Importantly, we highlighted that restarted and remnant radio sources are not merely on or off but comprise far more complex phases. These phases need to be further investigated to fully understand (1) the radio galaxies' life cycle, (2) the activity of their nuclear regions, and (3) the impact of these objects on their surroundings. In the future, large, multi-frequency, and multi-redshift surveys will be the key to enabling robust statistical studies of the radio morphologies of both extended and core emissions. These surveys will benefit from high-resolution radio observations, e.g. using very-long-baseline interferometry, which will be crucial to relate the history of nuclear radio variability to the optical nuclear emission line and continuum properties. This effort, in turn, will enable us to establish tight evolutionary models accurately depicting the life-cycle of radio galaxies.

\begin{acknowledgements}
This paper makes use of the data obtained with the International LOFAR Telescope (ILT) under project code LT$10\_012$. LOFAR, the Low Frequency Array designed and constructed by ASTRON (Netherlands Institute for Radio Astronomy), has facilities
in several countries, that are owned by various parties (each with their own funding sources), and that are collectively operated by the International LOFAR Telescope (ILT) foundation under a joint scientific policy.
The National Radio Astronomy Observatory is a facility of the National Science Foundation operated under cooperative agreement by Associated Universities, Inc.

This work also made use of NumPy, Matplotlib and TOPCAT.

MB acknowledges support from the agreement ASI-INAF n. 2017-14-H.O, from the PRIN MIUR 2017PH3WAT “Blackout”, from the ERC-Stg ``DRANOEL", no. 714245 and from the ERC-Stg ``MAGCOW", no. 714196.
\end{acknowledgements}

\bibliographystyle{aa}
\bibliography{long_baselines}

\begin{thebibliography}{82}
\expandafter\ifx\csname natexlab\endcsname\relax\def\natexlab#1{#1}\fi

\bibitem[{{Barthel} {et~al.}(1985){Barthel}, {Schilizzi}, {Miley}, {Jagers}, \&
  {Strom}}]{1985A&A...148..243Barthel}
{Barthel}, P.~D., {Schilizzi}, R.~T., {Miley}, G.~K., {Jagers}, W.~J., \&
  {Strom}, R.~G. 1985, \aap, 148, 243

\bibitem[{{Baum} {et~al.}(1990){Baum}, {O'Dea}, {Murphy}, \& {de
  Bruyn}}]{1990A&A...232...19Baum}
{Baum}, S.~A., {O'Dea}, C.~P., {Murphy}, D.~W., \& {de Bruyn}, A.~G. 1990,
  \aap, 232, 19

\bibitem[{{Becker} {et~al.}(1995){Becker}, {White}, \&
  {Helfand}}]{1995ApJ...450..559BeckerFIRST}
{Becker}, R.~H., {White}, R.~L., \& {Helfand}, D.~J. 1995, \apj, 450, 559

\bibitem[{{Best} {et~al.}(2005){Best}, {Kauffmann}, {Heckman}, \&
  {Ivezi{\'c}}}]{2005MNRAS.362....9Best}
{Best}, P.~N., {Kauffmann}, G., {Heckman}, T.~M., \& {Ivezi{\'c}}, {\v Z}.
  2005, \mnras, 362, 9

\bibitem[{{Bicknell} {et~al.}(1997){Bicknell}, {Dopita}, \&
  {O'Dea}}]{1997ApJ...485..112Bicknell}
{Bicknell}, G.~V., {Dopita}, M.~A., \& {O'Dea}, C. P.~O. 1997, \apj, 485, 112

\bibitem[{{Blandford} \& {K{\"o}nigl}(1979)}]{1979ApJ...232...34Blandford}
{Blandford}, R.~D. \& {K{\"o}nigl}, A. 1979, \apj, 232, 34

\bibitem[{{Blanton} {et~al.}(2017){Blanton}, {Bershady}, {Abolfathi},
  {Albareti}, {Allende Prieto}, {Almeida}, {Alonso-Garc{\'{\i}}a}, {Anders},
  {Anderson}, {Andrews}, \& et~al.}]{2017AJ....154...28Blanton_SDSS)}
{Blanton}, M.~R., {Bershady}, M.~A., {Abolfathi}, B., {et~al.} 2017, \aj, 154,
  28

\bibitem[{{Bridle} {et~al.}(1994){Bridle}, {Hough}, {Lonsdale}, {Burns}, \&
  {Laing}}]{1994AJ....108..766Bridle}
{Bridle}, A.~H., {Hough}, D.~H., {Lonsdale}, C.~J., {Burns}, J.~O., \& {Laing},
  R.~A. 1994, \aj, 108, 766

\bibitem[{{Brienza} {et~al.}(2017){Brienza}, {Godfrey}, {Morganti}, {Prandoni},
  {Harwood}, {Mahony}, {Hardcastle}, {Murgia}, {R{\"o}ttgering}, {Shimwell}, \&
  {Shulevski}}]{2017A&A...606A..98B}
{Brienza}, M., {Godfrey}, L., {Morganti}, R., {et~al.} 2017, \aap, 606, A98

\bibitem[{{Brienza} {et~al.}(2020){Brienza}, {Morganti}, {Harwood}, {Duchet},
  {Rajpurohit}, {Shulevski}, {Hardcastle}, {Mahatma}, {Godfrey}, {Prandoni},
  {Shimwell}, \& {Intema}}]{2020A&A...638A..29Brienza388}
{Brienza}, M., {Morganti}, R., {Harwood}, J., {et~al.} 2020, \aap, 638, A29

\bibitem[{{Brocksopp} {et~al.}(2011){Brocksopp}, {Kaiser}, {Schoenmakers}, \&
  {de Bruyn}}]{2011MNRAS.410..484Brocksopp_DDRG}
{Brocksopp}, C., {Kaiser}, C.~R., {Schoenmakers}, A.~P., \& {de Bruyn}, A.~G.
  2011, \mnras, 410, 484

\bibitem[{{Bruni} {et~al.}(2019){Bruni}, {Panessa}, {Bassani}, {Chiaraluce},
  {Kraus}, {Dallacasa}, {Bazzano}, {Hern{\'a}ndez-Garc{\'\i}a}, {Malizia},
  {Ubertini}, {Ursini}, \& {Venturi}}]{2019ApJ...875...88Bruni}
{Bruni}, G., {Panessa}, F., {Bassani}, L., {et~al.} 2019, \apj, 875, 88

\bibitem[{{Burns} {et~al.}(1982){Burns}, {Christiansen}, \&
  {Hough}}]{1982ApJ...257..538Burns}
{Burns}, J.~O., {Christiansen}, W.~A., \& {Hough}, D.~H. 1982, \apj, 257, 538

\bibitem[{{Callingham} {et~al.}(2017){Callingham}, {Ekers}, {Gaensler}, {Line},
  {Hurley-Walker}, {Sadler}, {Tingay}, {Hancock}, {Bell}, {Dwarakanath}, {For},
  {Franzen}, {Hindson}, {Johnston-Hollitt}, {Kapi{\'n}ska}, {Lenc}, {McKinley},
  {Morgan}, {Offringa}, {Procopio}, {Staveley-Smith}, {Wayth}, {Wu}, \&
  {Zheng}}]{2017ApJ...836..174Callingham}
{Callingham}, J.~R., {Ekers}, R.~D., {Gaensler}, B.~M., {et~al.} 2017, \apj,
  836, 174

\bibitem[{{Callingham} {et~al.}(2015){Callingham}, {Gaensler}, {Ekers},
  {Tingay}, {Wayth}, {Morgan}, {Bernardi}, {Bell}, {Bhat}, {Bowman}, {Briggs},
  {Cappallo}, {Deshpande}, {Ewall-Wice}, {Feng}, {Greenhill}, {Hazelton},
  {Hindson}, {Hurley-Walker}, {Jacobs}, {Johnston-Hollitt}, {Kaplan},
  {Kudrayvtseva}, {Lenc}, {Lonsdale}, {McKinley}, {McWhirter}, {Mitchell},
  {Morales}, {Morgan}, {Oberoi}, {Offringa}, {Ord}, {Pindor}, {Prabu},
  {Procopio}, {Riding}, {Srivani}, {Subrahmanyan}, {Udaya Shankar}, {Webster},
  {Williams}, \& {Williams}}]{2015ApJ...809..168Callingham}
{Callingham}, J.~R., {Gaensler}, B.~M., {Ekers}, R.~D., {et~al.} 2015, \apj,
  809, 168

\bibitem[{{Capetti} \& {Brienza}(2023)}]{2023A&A...676A.102CapettiBrienza}
{Capetti}, A. \& {Brienza}, M. 2023, \aap, 676, A102

\bibitem[{{Coppejans} {et~al.}(2016){Coppejans}, {Cseh}, {van Velzen},
  {Falcke}, {Intema}, {Paragi}, {M{\"u}ller}, {Williams}, {Frey}, {Gurvits}, \&
  {K{\"o}rding}}]{2016MNRAS.459.2455CoppejansMPS2}
{Coppejans}, R., {Cseh}, D., {van Velzen}, S., {et~al.} 2016, \mnras, 459, 2455

\bibitem[{{Coppejans} {et~al.}(2015){Coppejans}, {Cseh}, {Williams}, {van
  Velzen}, \& {Falcke}}]{2015MNRAS.450.1477CoppejansMPS1}
{Coppejans}, R., {Cseh}, D., {Williams}, W.~L., {van Velzen}, S., \& {Falcke},
  H. 2015, \mnras, 450, 1477

\bibitem[{{de Ruiter} {et~al.}(1990){de Ruiter}, {Parma}, {Fanti}, \&
  {Fanti}}]{1990A&A...227..351DeRuiter}
{de Ruiter}, H.~R., {Parma}, P., {Fanti}, C., \& {Fanti}, R. 1990, \aap, 227,
  351

\bibitem[{{Duncan} {et~al.}(2021){Duncan}, {Kondapally}, {Brown}, {Bonato},
  {Best}, {R{\"o}ttgering}, {Bondi}, {Bowler}, {Cochrane}, {G{\"u}rkan},
  {Hardcastle}, {Jarvis}, {Kunert-Bajraszewska}, {Leslie}, {Ma{\l}ek},
  {Morabito}, {O'Sullivan}, {Prandoni}, {Sabater}, {Shimwell}, {Smith}, {Wang},
  {Wo{\l}owska}, \& {Tasse}}]{2021A&A...648A...4Duncan}
{Duncan}, K.~J., {Kondapally}, R., {Brown}, M.~J.~I., {et~al.} 2021, \aap, 648,
  A4

\bibitem[{{Edwards} \& {Tingay}(2004)}]{2004A&A...424...91Edwards&Tingay}
{Edwards}, P.~G. \& {Tingay}, S.~J. 2004, \aap, 424, 91

\bibitem[{{Fanaroff} \& {Riley}(1974)}]{1974MNRAS.167P..31Fanaroff}
{Fanaroff}, B.~L. \& {Riley}, J.~M. 1974, \mnras, 167, 31P

\bibitem[{{Fanti}(2009)}]{2009AN....330..120Fanti}
{Fanti}, C. 2009, Astronomische Nachrichten, 330, 120

\bibitem[{{Fanti} {et~al.}(1990){Fanti}, {Fanti}, {Schilizzi}, {Spencer}, {Nan
  Rendong}, {Parma}, {van Breugel}, \& {Venturi}}]{1990A&A...231..333Fanti}
{Fanti}, R., {Fanti}, C., {Schilizzi}, R.~T., {et~al.} 1990, \aap, 231, 333

\bibitem[{{Feretti} {et~al.}(1984){Feretti}, {Giovannini}, {Gregorini},
  {Parma}, \& {Zamorani}}]{1984A&A...139...55Feretti}
{Feretti}, L., {Giovannini}, G., {Gregorini}, L., {Parma}, P., \& {Zamorani},
  G. 1984, \aap, 139, 55

\bibitem[{{Giovannini} {et~al.}(1988){Giovannini}, {Feretti}, {Gregorini}, \&
  {Parma}}]{1988A&A...199...73Giovannini}
{Giovannini}, G., {Feretti}, L., {Gregorini}, L., \& {Parma}, P. 1988, \aap,
  199, 73

\bibitem[{{Godfrey} {et~al.}(2017){Godfrey}, {Morganti}, \&
  {Brienza}}]{2017MNRAS.471..891Godfrey}
{Godfrey}, L.~E.~H., {Morganti}, R., \& {Brienza}, M. 2017, \mnras, 471, 891

\bibitem[{{Hancock} {et~al.}(2010){Hancock}, {Sadler}, {Mahony}, \&
  {Ricci}}]{2010MNRAS.408.1187Hancock}
{Hancock}, P.~J., {Sadler}, E.~M., {Mahony}, E.~K., \& {Ricci}, R. 2010,
  \mnras, 408, 1187

\bibitem[{{Hardcastle}(2018)}]{2018MNRAS.475.2768Hardcastle}
{Hardcastle}, M.~J. 2018, \mnras, 475, 2768

\bibitem[{{Jamrozy} {et~al.}(2009){Jamrozy}, {Konar}, {Saikia}, \&
  {Machalski}}]{2009ASPC..407..137Jamrozy_DDRG}
{Jamrozy}, M., {Konar}, C., {Saikia}, D.~J., \& {Machalski}, J. 2009, in
  Astronomical Society of the Pacific Conference Series, Vol. 407, The
  Low-Frequency Radio Universe, ed. D.~J. {Saikia}, D.~A. {Green}, Y.~{Gupta},
  \& T.~{Venturi}, 137

\bibitem[{{Jurlin} {et~al.}(2021{\natexlab{a}}){Jurlin}, {Brienza}, {Morganti},
  {Wadadekar}, {Ishwara-Chandra}, {Maddox}, \&
  {Mahatma}}]{2021A&A...653A.110Jurlin}
{Jurlin}, N., {Brienza}, M., {Morganti}, R., {et~al.} 2021{\natexlab{a}}, \aap,
  653, A110

\bibitem[{{Jurlin} {et~al.}(2020){Jurlin}, {Morganti}, {Brienza}, {Mandal},
  {Maddox}, {Duncan}, {Shabala}, {Hardcastle}, {Prandoni}, {R{\"o}ttgering},
  {Mahatma}, {Best}, {Mingo}, {Sabater}, {Shimwell}, \&
  {Tasse}}]{2020A&A...638A..34Jurlin}
{Jurlin}, N., {Morganti}, R., {Brienza}, M., {et~al.} 2020, \aap, 638, A34

\bibitem[{{Jurlin} {et~al.}(2021{\natexlab{b}}){Jurlin}, {Morganti}, {Maddox},
  \& {Brienza}}]{2021Galax...9..122Jurlin}
{Jurlin}, N., {Morganti}, R., {Maddox}, N., \& {Brienza}, M.
  2021{\natexlab{b}}, Galaxies, 9, 122

\bibitem[{{Kaiser} {et~al.}(2000){Kaiser}, {Schoenmakers}, \&
  {R{\"o}ttgering}}]{2000MNRAS.315..381Kaiser_DDRG}
{Kaiser}, C.~R., {Schoenmakers}, A.~P., \& {R{\"o}ttgering}, H. J.~A. 2000,
  \mnras, 315, 381

\bibitem[{{Kesteven} {et~al.}(1977){Kesteven}, {Bridle}, \&
  {Brandie}}]{1977AJ.....82..541Kesteven}
{Kesteven}, M.~J.~L., {Bridle}, A.~H., \& {Brandie}, G.~W. 1977, \aj, 82, 541

\bibitem[{Kolmogorov(1933)}]{1933Kolmogorov}
Kolmogorov, A. 1933, G. Ist. Ital. Attuari., 4, 83

\bibitem[{{Konar} {et~al.}(2012){Konar}, {Hardcastle}, {Jamrozy}, {Croston}, \&
  {Nandi}}]{2012MNRAS.424.1061Konar_DDRG}
{Konar}, C., {Hardcastle}, M.~J., {Jamrozy}, M., {Croston}, J.~H., \& {Nandi},
  S. 2012, \mnras, 424, 1061

\bibitem[{{Kondapally} {et~al.}(2021){Kondapally}, {Best}, {Hardcastle},
  {Nisbet}, {Bonato}, {Sabater}, {Duncan}, {McCheyne}, {Cochrane}, {Bowler},
  {Williams}, {Shimwell}, {Tasse}, {Croston}, {Goyal}, {Jamrozy}, {Jarvis},
  {Mahatma}, {R{\"o}ttgering}, {Smith}, {Wo{\l}owska}, {Bondi}, {Brienza},
  {Brown}, {Br{\"u}ggen}, {Chambers}, {Garrett}, {G{\"u}rkan}, {Huber},
  {Kunert-Bajraszewska}, {Magnier}, {Mingo}, {Mostert},
  {Nikiel-Wroczy{\'n}ski}, {O'Sullivan}, {Paladino}, {Ploeckinger}, {Prandoni},
  {Rosenthal}, {Schwarz}, {Shulevski}, {Wagenveld}, \&
  {Wang}}]{2021A&A...648A...3Kondapally}
{Kondapally}, R., {Best}, P.~N., {Hardcastle}, M.~J., {et~al.} 2021, \aap, 648,
  A3

\bibitem[{{Lacy} {et~al.}(2020){Lacy}, {Baum}, {Chandler}, {Chatterjee},
  {Clarke}, {Deustua}, {English}, {Farnes}, {Gaensler}, {Gugliucci},
  {Hallinan}, {Kent}, {Kimball}, {Law}, {Lazio}, {Marvil}, {Mao}, {Medlin},
  {Mooley}, {Murphy}, {Myers}, {Osten}, {Richards}, {Rosolowsky}, {Rudnick},
  {Schinzel}, {Sivakoff}, {Sjouwerman}, {Taylor}, {White}, {Wrobel},
  {Andernach}, {Beasley}, {Berger}, {Bhatnager}, {Birkinshaw}, {Bower},
  {Brandt}, {Brown}, {Burke-Spolaor}, {Butler}, {Comerford}, {Demorest}, {Fu},
  {Giacintucci}, {Golap}, {G{\"u}th}, {Hales}, {Hiriart}, {Hodge}, {Horesh},
  {Ivezi{\'c}}, {Jarvis}, {Kamble}, {Kassim}, {Liu}, {Loinard}, {Lyons},
  {Masters}, {Mezcua}, {Moellenbrock}, {Mroczkowski}, {Nyland},
  {O{\textquoteright}Dea}, {O{\textquoteright}Sullivan}, {Peters}, {Radford},
  {Rao}, {Robnett}, {Salcido}, {Shen}, {Sobotka}, {Witz}, {Vaccari}, {van
  Weeren}, {Vargas}, {Williams}, \& {Yoon}}]{2020PASP..132c5001LacyVLASS}
{Lacy}, M., {Baum}, S.~A., {Chandler}, C.~J., {et~al.} 2020, \pasp, 132, 035001

\bibitem[{{Lockman} {et~al.}(1986){Lockman}, {Jahoda}, \&
  {McCammon}}]{1986ApJ...302..432Lockman_LH}
{Lockman}, F.~J., {Jahoda}, K., \& {McCammon}, D. 1986, \apj, 302, 432

\bibitem[{{Mahatma} {et~al.}(2019){Mahatma}, {Hardcastle}, {Williams}, {Best},
  {Croston}, {Duncan}, {Mingo}, {Morganti}, {Brienza}, {Cochrane},
  {G{\"u}rkan}, {Harwood}, {Jarvis}, {Jamrozy}, {Jurlin}, {Morabito},
  {R{\"o}ttgering}, {Sabater}, {Shimwell}, {Smith}, {Shulevski}, \&
  {Tasse}}]{2019A&A...622A..13Mahatma_restarted}
{Mahatma}, V.~H., {Hardcastle}, M.~J., {Williams}, W.~L., {et~al.} 2019, \aap,
  622, A13

\bibitem[{{Mahony} {et~al.}(2016){Mahony}, {Morganti}, {Prandoni}, {van
  Bemmel}, {Shimwell}, {Brienza}, {Best}, {Br{\"u}ggen}, {Calistro Rivera}, {de
  Gasperin}, {Hardcastle}, {Harwood}, {Heald}, {Jarvis}, {Mandal}, {Miley},
  {Retana-Montenegro}, {R{\"o}ttgering}, {Sabater}, {Tasse}, {van Velzen}, {van
  Weeren}, {Williams}, \& {White}}]{2016MNRAS.463.2997Mahony}
{Mahony}, E.~K., {Morganti}, R., {Prandoni}, I., {et~al.} 2016, \mnras, 463,
  2997

\bibitem[{{Massardi} {et~al.}(2011){Massardi}, {Ekers}, {Murphy}, {Mahony},
  {Hancock}, {Chhetri}, {de Zotti}, {Sadler}, {Burke-Spolaor}, {Calabretta},
  {Edwards}, {Ekers}, {Jackson}, {Kesteven}, {Newton-McGee}, {Phillips},
  {Ricci}, {Roberts}, {Sault}, {Staveley-Smith}, {Subrahmanyan}, {Walker}, \&
  {Wilson}}]{2011MNRAS.412..318Massardi}
{Massardi}, M., {Ekers}, R.~D., {Murphy}, T., {et~al.} 2011, \mnras, 412, 318

\bibitem[{{McCaffrey} {et~al.}(2022){McCaffrey}, {Kimball}, {Momjian}, \&
  {Richards}}]{2022AJ....164..122McCaffrey}
{McCaffrey}, T.~V., {Kimball}, A.~E., {Momjian}, E., \& {Richards}, G.~T. 2022,
  \aj, 164, 122

\bibitem[{{Mingo} {et~al.}(2019){Mingo}, {Croston}, {Hardcastle}, {Best},
  {Duncan}, {Morganti}, {Rottgering}, {Sabater}, {Shimwell}, {Williams},
  {Brienza}, {Gurkan}, {Mahatma}, {Morabito}, {Prandoni}, {Bondi}, {Ineson}, \&
  {Mooney}}]{2019MNRAS.488.2701Mingo}
{Mingo}, B., {Croston}, J.~H., {Hardcastle}, M.~J., {et~al.} 2019, \mnras, 488,
  2701

\bibitem[{{Morabito} {et~al.}(2022){Morabito}, {Jackson}, {Mooney}, {Sweijen},
  {Badole}, {Kukreti}, {Venkattu}, {Groeneveld}, {Kappes}, {Bonnassieux},
  {Drabent}, {Iacobelli}, {Croston}, {Best}, {Bondi}, {Callingham}, {Conway},
  {Deller}, {Hardcastle}, {McKean}, {Miley}, {Moldon}, {R{\"o}ttgering},
  {Tasse}, {Shimwell}, {van Weeren}, {Anderson}, {Asgekar}, {Avruch}, {van
  Bemmel}, {Bentum}, {Bonafede}, {Brouw}, {Butcher}, {Ciardi}, {Corstanje},
  {Coolen}, {Damstra}, {de Gasperin}, {Duscha}, {Eisl{\"o}ffel}, {Engels},
  {Falcke}, {Garrett}, {Griessmeier}, {Gunst}, {van Haarlem}, {Hoeft}, {van der
  Horst}, {J{\"u}tte}, {Kadler}, {Koopmans}, {Krankowski}, {Mann}, {Nelles},
  {Oonk}, {Orru}, {Paas}, {Pandey}, {Pizzo}, {Pandey-Pommier}, {Reich},
  {Rothkaehl}, {Ruiter}, {Schwarz}, {Shulevski}, {Soida}, {Tagger}, {Vocks},
  {Wijers}, {Wijnholds}, {Wucknitz}, {Zarka}, \&
  {Zucca}}]{2022A&A...658A...1Morabito}
{Morabito}, L.~K., {Jackson}, N.~J., {Mooney}, S., {et~al.} 2022, \aap, 658, A1

\bibitem[{{Morganti} {et~al.}(2021{\natexlab{a}}){Morganti}, {Jurlin},
  {Oosterloo}, {Brienza}, {Orr{\'u}}, {Kutkin}, {Prandoni}, {Adams},
  {D{\'e}nes}, {Hess}, {Shulevski}, {van der Hulst}, \&
  {Ziemke}}]{2021Galax...9...88Morganti_resolved_si2}
{Morganti}, R., {Jurlin}, N., {Oosterloo}, T., {et~al.} 2021{\natexlab{a}},
  Galaxies, 9, 88

\bibitem[{{Morganti} {et~al.}(2021{\natexlab{b}}){Morganti}, {Oosterloo},
  {Brienza}, {Jurlin}, {Prandoni}, {Orr{\`u}}, {Shabala}, {Adams}, {Adebahr},
  {Best}, {Coolen}, {Damstra}, {de Blok}, {de Gasperin}, {D{\'e}nes},
  {Hardcastle}, {Hess}, {Hut}, {Kondapally}, {Kutkin}, {Loose}, {Lucero},
  {Maan}, {Maccagni}, {Mingo}, {Moss}, {Mostert}, {Norden}, {Oostrum},
  {R{\"o}ttgering}, {Ruiter}, {Shimwell}, {Schulz}, {Vermaas}, {Vohl}, {van der
  Hulst}, {van Diepen}, {van Leeuwen}, \&
  {Ziemke}}]{2021A&A...648A...9Morganti_resolvedsi}
{Morganti}, R., {Oosterloo}, T.~A., {Brienza}, M., {et~al.} 2021{\natexlab{b}},
  \aap, 648, A9

\bibitem[{{Mullin} {et~al.}(2008){Mullin}, {Riley}, \&
  {Hardcastle}}]{2008MNRAS.390..595Mullin}
{Mullin}, L.~M., {Riley}, J.~M., \& {Hardcastle}, M.~J. 2008, \mnras, 390, 595

\bibitem[{{Nyland} {et~al.}(2020){Nyland}, {Dong}, {Patil}, {Lacy}, {van
  Velzen}, {Kimball}, {Sarbadhicary}, {Hallinan}, {Baldassare}, {Clarke},
  {Goulding}, {Greene}, {Hughes}, {Kassim}, {Kunert-Bajraszewska}, {Maccarone},
  {Mooley}, {Mukherjee}, {Peters}, {Petrov}, {Polisensky}, {Rujopakarn},
  {Whittle}, \& {Vaccari}}]{2020ApJ...905...74NylandVARIABILITY}
{Nyland}, K., {Dong}, D.~Z., {Patil}, P., {et~al.} 2020, \apj, 905, 74

\bibitem[{{O'Dea}(1998)}]{1998PASP..110..493Odea}
{O'Dea}, C.~P. 1998, \pasp, 110, 493

\bibitem[{{O'Dea} \& {Baum}(1997)}]{1997AJ....113..148O'Dea_Baum}
{O'Dea}, C.~P. \& {Baum}, S.~A. 1997, \aj, 113, 148

\bibitem[{{O'Dea} \& {Saikia}(2021)}]{2021A&ARv..29....3OdeaSaikia_review}
{O'Dea}, C.~P. \& {Saikia}, D.~J. 2021, \aapr, 29, 3

\bibitem[{{Orienti} \& {Dallacasa}(2008)}]{2008A&A...487..885Orienti_Dallacasa}
{Orienti}, M. \& {Dallacasa}, D. 2008, \aap, 487, 885

\bibitem[{{Parma} {et~al.}(1987){Parma}, {Fanti}, {Fanti}, {Morganti}, \& {de
  Ruiter}}]{1987A&A...181..244Parma}
{Parma}, P., {Fanti}, C., {Fanti}, R., {Morganti}, R., \& {de Ruiter}, H.~R.
  1987, \aap, 181, 244

\bibitem[{{Parma} {et~al.}(2010){Parma}, {Mantovani}, {de Ruiter}, {Mack},
  {Murgia}, \& {Govoni}}]{2010evn..confE..89ParmaVLBI}
{Parma}, P., {Mantovani}, F., {de Ruiter}, H.~R., {et~al.} 2010, in 10th
  European VLBI Network Symposium and EVN Users Meeting: VLBI and the New
  Generation of Radio Arrays, Vol.~10, 89

\bibitem[{{Parma} {et~al.}(1999){Parma}, {Murgia}, {Morganti}, {Capetti}, {de
  Ruiter}, \& {Fanti}}]{1999A&A...344....7Parma}
{Parma}, P., {Murgia}, M., {Morganti}, R., {et~al.} 1999, \aap, 344, 7

\bibitem[{{Patil} {et~al.}(2022){Patil}, {Whittle}, {Nyland}, {Lonsdale},
  {Lacy}, {Kimball}, {Lonsdale}, {Peters}, {Clarke}, {Efstathiou},
  {Giacintucci}, {Kim}, {Lanz}, {Mukherjee}, \&
  {Polisensky}}]{2022ApJ...934...26Patil_cc_plot}
{Patil}, P., {Whittle}, M., {Nyland}, K., {et~al.} 2022, \apj, 934, 26

\bibitem[{{Peck} {et~al.}(1999){Peck}, {Taylor}, \&
  {Conway}}]{1999ApJ...521..103Peck}
{Peck}, A.~B., {Taylor}, G.~B., \& {Conway}, J.~E. 1999, \apj, 521, 103

\bibitem[{{Perley} \& {Butler}(2017)}]{2017ApJS..230....7Perley}
{Perley}, R.~A. \& {Butler}, B.~J. 2017, \apjs, 230, 7

\bibitem[{{Roettiger} {et~al.}(1994){Roettiger}, {Burns}, {Clarke}, \&
  {Christiansen}}]{1994ApJ...421L..23Roettiger}
{Roettiger}, K., {Burns}, J.~O., {Clarke}, D.~A., \& {Christiansen}, W.~A.
  1994, \apjl, 421, L23

\bibitem[{{Ross} {et~al.}(2021){Ross}, {Callingham}, {Hurley-Walker},
  {Seymour}, {Hancock}, {Franzen}, {Morgan}, {White}, {Bell}, \&
  {Patil}}]{2021MNRAS.501.6139Ross_spectral_variability}
{Ross}, K., {Callingham}, J.~R., {Hurley-Walker}, N., {et~al.} 2021, \mnras,
  501, 6139

\bibitem[{{Rudnick} {et~al.}(1986){Rudnick}, {Jones}, \&
  {Fiedler}}]{1986AJ.....91.1011Rudnick}
{Rudnick}, L., {Jones}, T.~W., \& {Fiedler}, R. 1986, \aj, 91, 1011

\bibitem[{{Sabater} {et~al.}(2019){Sabater}, {Best}, {Hardcastle}, {Shimwell},
  {Tasse}, {Williams}, {Br{\"u}ggen}, {Cochrane}, {Croston}, {de Gasperin},
  {Duncan}, {G{\"u}rkan}, {Mechev}, {Morabito}, {Prandoni}, {R{\"o}ttgering},
  {Smith}, {Harwood}, {Mingo}, {Mooney}, \&
  {Saxena}}]{2019A&A...622A..17Sabater}
{Sabater}, J., {Best}, P.~N., {Hardcastle}, M.~J., {et~al.} 2019, \aap, 622,
  A17

\bibitem[{{Sadler} {et~al.}(2006){Sadler}, {Ricci}, {Ekers}, {Ekers},
  {Hancock}, {Jackson}, {Kesteven}, {Murphy}, {Phillips}, {Reinfrank},
  {Staveley-Smith}, {Subrahmanyan}, {Walker}, {Wilson}, \& {de
  Zotti}}]{2006MNRAS.371..898Sadler}
{Sadler}, E.~M., {Ricci}, R., {Ekers}, R.~D., {et~al.} 2006, \mnras, 371, 898

\bibitem[{{Saikia} {et~al.}(2006){Saikia}, {Konar}, \&
  {Kulkarni}}]{2006MNRAS.366.1391Saikia_DDRG}
{Saikia}, D.~J., {Konar}, C., \& {Kulkarni}, V.~K. 2006, \mnras, 366, 1391

\bibitem[{{Saripalli} {et~al.}(2012){Saripalli}, {Subrahmanyan}, {Thorat},
  {Ekers}, {Hunstead}, {Johnston}, \& {Sadler}}]{2012ApJS..199...27Saripalli}
{Saripalli}, L., {Subrahmanyan}, R., {Thorat}, K., {et~al.} 2012, \apjs, 199,
  27

\bibitem[{{Saripalli} {et~al.}(2003){Saripalli}, {Subrahmanyan}, \& {Udaya
  Shankar}}]{2003ApJ...590..181Saripalli_DDRG}
{Saripalli}, L., {Subrahmanyan}, R., \& {Udaya Shankar}, N. 2003, \apj, 590,
  181

\bibitem[{{Schoenmakers} {et~al.}(2000){Schoenmakers}, {de Bruyn},
  {R{\"o}ttgering}, {van der Laan}, \&
  {Kaiser}}]{2000MNRAS.315..371Schoenmakers}
{Schoenmakers}, A.~P., {de Bruyn}, A.~G., {R{\"o}ttgering}, H.~J.~A., {van der
  Laan}, H., \& {Kaiser}, C.~R. 2000, \mnras, 315, 371

\bibitem[{{Shabala} {et~al.}(2020){Shabala}, {Jurlin}, {Morganti}, {Brienza},
  {Hardcastle}, {Godfrey}, {Krause}, \& {Turner}}]{2020MNRAS.496.1706Shabala}
{Shabala}, S.~S., {Jurlin}, N., {Morganti}, R., {et~al.} 2020, \mnras, 496,
  1706

\bibitem[{{Shimwell} {et~al.}(2019){Shimwell}, {Tasse}, {Hardcastle}, {Mechev},
  {Williams}, {Best}, {R{\"o}ttgering}, {Callingham}, {Dijkema}, {de Gasperin},
  {Hoang}, {Hugo}, {Mirmont}, {Oonk}, {Prandoni}, {Rafferty}, {Sabater},
  {Smirnov}, {van Weeren}, {White}, {Atemkeng}, {Bester}, {Bonnassieux},
  {Br{\"u}ggen}, {Brunetti}, {Chy{\.z}y}, {Cochrane}, {Conway}, {Croston},
  {Danezi}, {Duncan}, {Haverkorn}, {Heald}, {Iacobelli}, {Intema}, {Jackson},
  {Jamrozy}, {Jarvis}, {Lakhoo}, {Mevius}, {Miley}, {Morabito}, {Morganti},
  {Nisbet}, {Orr{\'u}}, {Perkins}, {Pizzo}, {Schrijvers}, {Smith}, {Vermeulen},
  {Wise}, {Alegre}, {Bacon}, {van Bemmel}, {Beswick}, {Bonafede}, {Botteon},
  {Bourke}, {Brienza}, {Calistro Rivera}, {Cassano}, {Clarke}, {Conselice},
  {Dettmar}, {Drabent}, {Dumba}, {Emig}, {En{\ss}lin}, {Ferrari}, {Garrett},
  {G{\'e}nova-Santos}, {Goyal}, {G{\"u}rkan}, {Hale}, {Harwood}, {Heesen},
  {Hoeft}, {Horellou}, {Jackson}, {Kokotanekov}, {Kondapally},
  {Kunert-Bajraszewska}, {Mahatma}, {Mahony}, {Mandal}, {McKean}, {Merloni},
  {Mingo}, {Miskolczi}, {Mooney}, {Nikiel-Wroczy{\'n}ski}, {O'Sullivan},
  {Quinn}, {Reich}, {Roskowi{\'n}ski}, {Rowlinson}, {Savini}, {Saxena},
  {Schwarz}, {Shulevski}, {Sridhar}, {Stacey}, {Urquhart}, {van der Wiel},
  {Varenius}, {Webster}, \& {Wilber}}]{2019A&A...622A...1Shimwell}
{Shimwell}, T.~W., {Tasse}, C., {Hardcastle}, M.~J., {et~al.} 2019, \aap, 622,
  A1

\bibitem[{{Shulevski} {et~al.}(2012){Shulevski}, {Morganti}, {Oosterloo}, \&
  {Struve}}]{2012A&A...545A..91Shulevski}
{Shulevski}, A., {Morganti}, R., {Oosterloo}, T., \& {Struve}, C. 2012, \aap,
  545, A91

\bibitem[{{Stanghellini} {et~al.}(2005){Stanghellini}, {O'Dea}, {Dallacasa},
  {Cassaro}, {Baum}, {Fanti}, \& {Fanti}}]{2005A&A...443..891Stanghellini}
{Stanghellini}, C., {O'Dea}, C.~P., {Dallacasa}, D., {et~al.} 2005, \aap, 443,
  891

\bibitem[{{Subrahmanyan} {et~al.}(2010){Subrahmanyan}, {Ekers}, {Saripalli}, \&
  {Sadler}}]{2010MNRAS.402.2792Subrahmanyan_ATLBS}
{Subrahmanyan}, R., {Ekers}, R.~D., {Saripalli}, L., \& {Sadler}, E.~M. 2010,
  \mnras, 402, 2792

\bibitem[{{Sweijen} {et~al.}(2023){Sweijen}, {Lyu}, {Wang}, {Gao},
  {R{\"o}ttgering}, {van Weeren}, {Morabito}, {Best}, {Ma{\l}ek}, {Williams},
  {Prandoni}, {Bonato}, \& {Bondi}}]{2023A&A...671A..85Sweijen}
{Sweijen}, F., {Lyu}, Y., {Wang}, L., {et~al.} 2023, \aap, 671, A85

\bibitem[{{Sweijen} {et~al.}(2022){Sweijen}, {van Weeren}, {R{\"o}ttgering},
  {Morabito}, {Jackson}, {Offringa}, {van der Tol}, {Veenboer}, {Oonk}, {Best},
  {Bondi}, {Shimwell}, {Tasse}, \& {Thomson}}]{2022NatAs.tmp...24Sweijen}
{Sweijen}, F., {van Weeren}, R.~J., {R{\"o}ttgering}, H.~J.~A., {et~al.} 2022,
  Nature Astronomy [\eprint[arXiv]{2202.01608}]

\bibitem[{{Tasse} {et~al.}(2021){Tasse}, {Shimwell}, {Hardcastle},
  {O'Sullivan}, {van Weeren}, {Best}, {Bester}, {Hugo}, {Smirnov}, {Sabater},
  {Calistro-Rivera}, {de Gasperin}, {Morabito}, {R{\"o}ttgering}, {Williams},
  {Bonato}, {Bondi}, {Botteon}, {Br{\"u}ggen}, {Brunetti}, {Chy{\.z}y},
  {Garrett}, {G{\"u}rkan}, {Jarvis}, {Kondapally}, {Mandal}, {Prandoni},
  {Repetti}, {Retana-Montenegro}, {Schwarz}, {Shulevski}, \&
  {Wiaux}}]{2021A&A...648A...1Tasse_DEEP_FIELDS}
{Tasse}, C., {Shimwell}, T., {Hardcastle}, M.~J., {et~al.} 2021, \aap, 648, A1

\bibitem[{{Tingay} {et~al.}(2015){Tingay}, {Macquart}, {Collier}, {Rees},
  {Callingham}, {Stevens}, {Carretti}, {Wayth}, {Wong}, {Trott}, {McKinley},
  {Bernardi}, {Bowman}, {Briggs}, {Cappallo}, {Corey}, {Deshpande}, {Emrich},
  {Gaensler}, {Goeke}, {Greenhill}, {Hazelton}, {Johnston-Hollitt}, {Kaplan},
  {Kasper}, {Kratzenberg}, {Lonsdale}, {Lynch}, {McWhirter}, {Mitchell},
  {Morales}, {Morgan}, {Oberoi}, {Ord}, {Prabu}, {Rogers}, {Roshi}, {Udaya
  Shankar}, {Srivani}, {Subrahmanyan}, {Waterson}, {Webster}, {Whitney},
  {Williams}, \& {Williams}}]{2015AJ....149...74Tingay}
{Tingay}, S.~J., {Macquart}, J.~P., {Collier}, J.~D., {et~al.} 2015, \aj, 149,
  74

\bibitem[{{Torniainen} {et~al.}(2007){Torniainen}, {Tornikoski},
  {L{\"a}hteenm{\"a}ki}, {Aller}, {Aller}, \&
  {Mingaliev}}]{2007A&A...469..451Torniainen}
{Torniainen}, I., {Tornikoski}, M., {L{\"a}hteenm{\"a}ki}, A., {et~al.} 2007,
  \aap, 469, 451

\bibitem[{{Tremblay} {et~al.}(2010){Tremblay}, {O'Dea}, {Baum}, {Koekemoer},
  {Sparks}, {de Bruyn}, \& {Schoenmakers}}]{2010ApJ...715..172Tremblay}
{Tremblay}, G.~R., {O'Dea}, C.~P., {Baum}, S.~A., {et~al.} 2010, \apj, 715, 172

\bibitem[{{van Haarlem} {et~al.}(2013){van Haarlem}, {Wise}, {Gunst}, {Heald},
  {McKean}, {Hessels}, {de Bruyn}, {Nijboer}, {Swinbank}, {Fallows},
  {Brentjens}, {Nelles}, {Beck}, {Falcke}, {Fender}, {H{\"o}randel},
  {Koopmans}, {Mann}, {Miley}, {R{\"o}ttgering}, {Stappers}, {Wijers},
  {Zaroubi}, {van den Akker}, {Alexov}, {Anderson}, {Anderson}, {van Ardenne},
  {Arts}, {Asgekar}, {Avruch}, {Batejat}, {B{\"a}hren}, {Bell}, {Bell}, {van
  Bemmel}, {Bennema}, {Bentum}, {Bernardi}, {Best}, {B{\^i}rzan}, {Bonafede},
  {Boonstra}, {Braun}, {Bregman}, {Breitling}, {van de Brink}, {Broderick},
  {Broekema}, {Brouw}, {Br{\"u}ggen}, {Butcher}, {van Cappellen}, {Ciardi},
  {Coenen}, {Conway}, {Coolen}, {Corstanje}, {Damstra}, {Davies}, {Deller},
  {Dettmar}, {van Diepen}, {Dijkstra}, {Donker}, {Doorduin}, {Dromer}, {Drost},
  {van Duin}, {Eisl{\"o}ffel}, {van Enst}, {Ferrari}, {Frieswijk}, {Gankema},
  {Garrett}, {de Gasperin}, {Gerbers}, {de Geus}, {Grie{\ss}meier}, {Grit},
  {Gruppen}, {Hamaker}, {Hassall}, {Hoeft}, {Holties}, {Horneffer}, {van der
  Horst}, {van Houwelingen}, {Huijgen}, {Iacobelli}, {Intema}, {Jackson},
  {Jelic}, {de Jong}, {Juette}, {Kant}, {Karastergiou}, {Koers}, {Kollen},
  {Kondratiev}, {Kooistra}, {Koopman}, {Koster}, {Kuniyoshi}, {Kramer},
  {Kuper}, {Lambropoulos}, {Law}, {van Leeuwen}, {Lemaitre}, {Loose}, {Maat},
  {Macario}, {Markoff}, {Masters}, {McFadden}, {McKay-Bukowski}, {Meijering},
  {Meulman}, {Mevius}, {Middelberg}, {Millenaar}, {Miller-Jones}, {Mohan},
  {Mol}, {Morawietz}, {Morganti}, {Mulcahy}, {Mulder}, {Munk}, {Nieuwenhuis},
  {van Nieuwpoort}, {Noordam}, {Norden}, {Noutsos}, {Offringa}, {Olofsson},
  {Omar}, {Orr{\'u}}, {Overeem}, {Paas}, {Pandey-Pommier}, {Pandey}, {Pizzo},
  {Polatidis}, {Rafferty}, {Rawlings}, {Reich}, {de Reijer}, {Reitsma},
  {Renting}, {Riemers}, {Rol}, {Romein}, {Roosjen}, {Ruiter}, {Scaife}, {van
  der Schaaf}, {Scheers}, {Schellart}, {Schoenmakers}, {Schoonderbeek},
  {Serylak}, {Shulevski}, {Sluman}, {Smirnov}, {Sobey}, {Spreeuw}, {Steinmetz},
  {Sterks}, {Stiepel}, {Stuurwold}, {Tagger}, {Tang}, {Tasse}, {Thomas},
  {Thoudam}, {Toribio}, {van der Tol}, {Usov}, {van Veelen}, {van der Veen},
  {ter Veen}, {Verbiest}, {Vermeulen}, {Vermaas}, {Vocks}, {Vogt}, {de Vos},
  {van der Wal}, {van Weeren}, {Weggemans}, {Weltevrede}, {White}, {Wijnholds},
  {Wilhelmsson}, {Wucknitz}, {Yatawatta}, {Zarka}, {Zensus}, \& {van
  Zwieten}}]{2013A&A...556A...2VanHaarlem}
{van Haarlem}, M.~P., {Wise}, M.~W., {Gunst}, A.~W., {et~al.} 2013, \aap, 556,
  A2

\bibitem[{{Willis} {et~al.}(1974){Willis}, {Strom}, \&
  {Wilson}}]{1974Natur.250..625Willis}
{Willis}, A.~G., {Strom}, R.~G., \& {Wilson}, A.~S. 1974, \nat, 250, 625

\end{thebibliography}

\appendix
\onecolumn
\section{Figures of sources studied in this work}
\begin{figure*} [!htp]
\minipage{\textwidth}
\hspace{0.45cm}
        \includegraphics[width=0.39\linewidth] {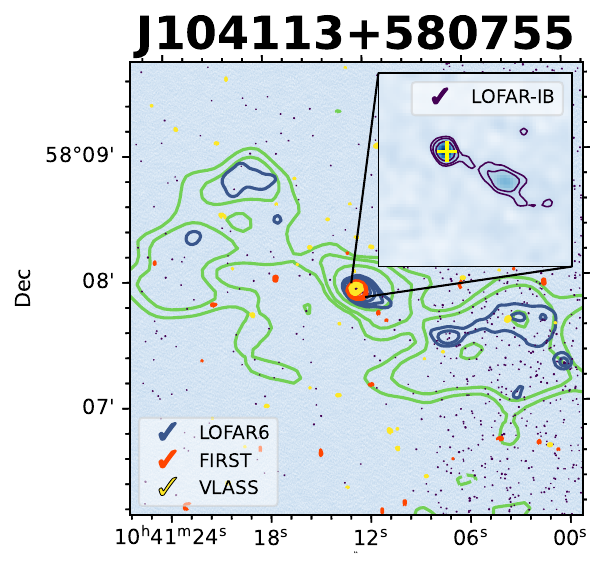} 
        \includegraphics[width=0.41\linewidth] {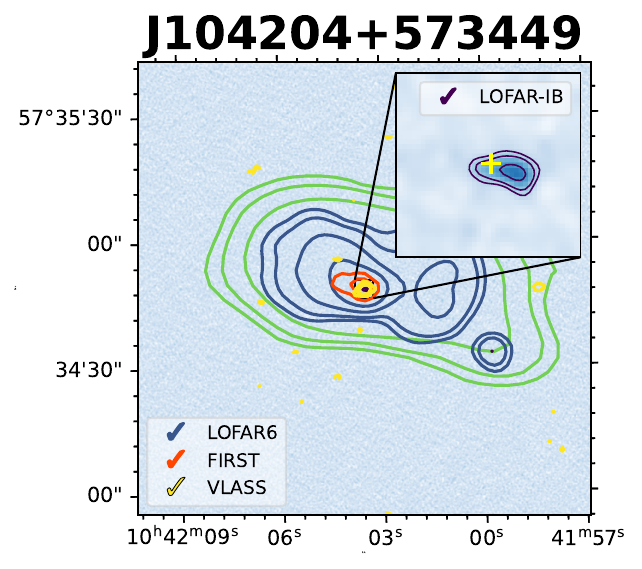}
\endminipage \hfill
\minipage{\textwidth}
        \includegraphics[width=0.42\linewidth] {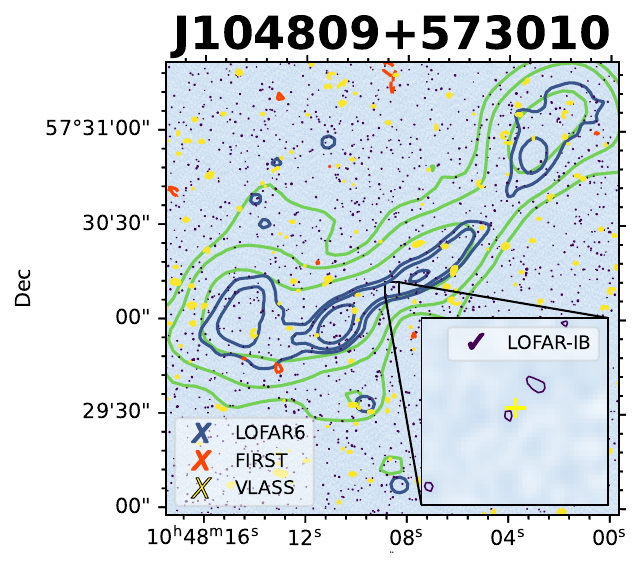}
        \includegraphics[width=0.40\linewidth] {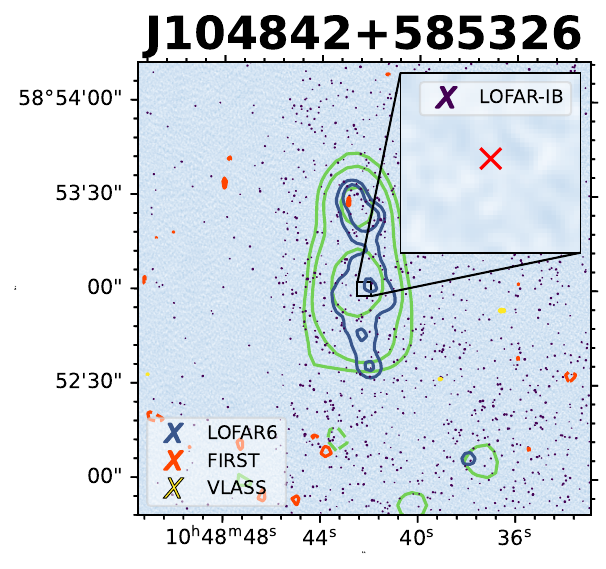}
\endminipage \hfill
\minipage{\textwidth}
        \includegraphics[width=0.41\linewidth] {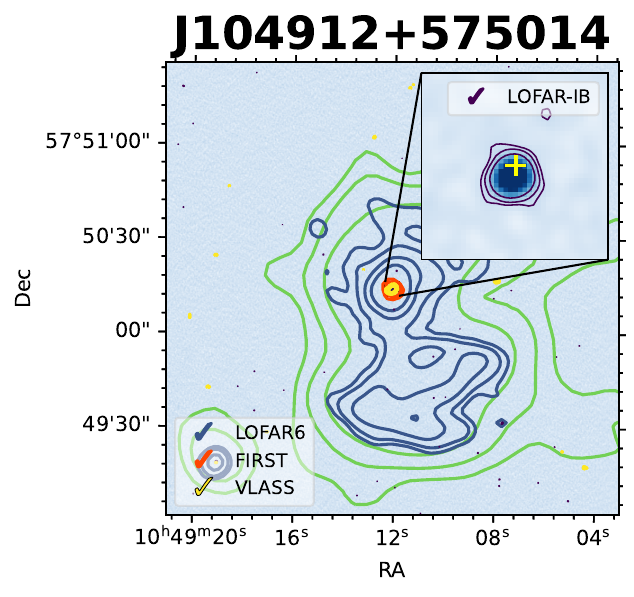}
        \includegraphics[width=0.40\linewidth] {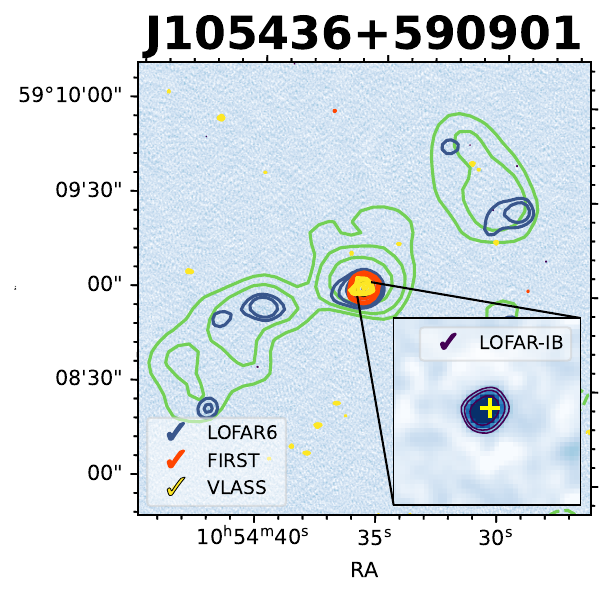}
\endminipage \hfill
\caption{Radio images of restarted candidates. The colourmaps are LOFAR-IB with contours from the LOFAR18 (green; -3, 3, 5, 10 $\times$ $\sigma_{\rm local}$); LOFAR6 (blue; -3, 3, 5, 10, 20, 100, 200 $\times$ $\sigma_{\rm median}$); LOFAR-IB (purple; -3, 3, 5, 10 $\times$ $\sigma_{\rm local}$); FIRST (red; -3, 3, 5, 10 $\times$ $\sigma_{\rm local}$); VLASS (yellow; -3, 3, 5, 10, 20 $\times$ $\sigma_{\rm local}$). The inset is zoomed in on the central region. The yellow `+' in the inset notes the position of the optical host galaxy. If there is no optical counterpart, red `$\times$' notes the expected position of the nuclear region. Detection (`\cmark') or non-detection (`\xmark') of the central region is noted in the legends.}
\label{fig:restarted}
\end{figure*}

\begin{figure*} [h]
\minipage{\textwidth}
\centering
        \includegraphics[width=0.48\linewidth] {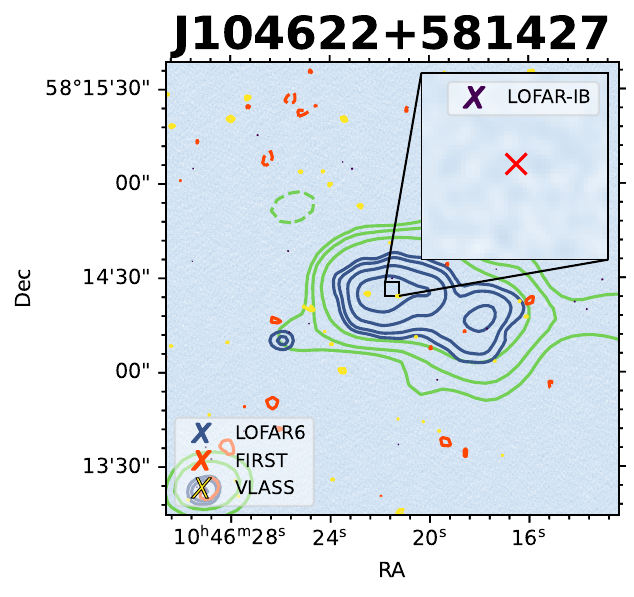}
\endminipage \hfill
\caption{Radio image of a remnant radio source. The colourmap is LOFAR-IB, with contours from the LOFAR18 image in green, LOFAR6 in blue, FIRST in red (no detection in the full image), VLASS in yellow, and LOFAR-IB in purple (no detection in the full extent of the source). Contour levels are the same as in Fig.~\ref{fig:restarted}. The inset is showing zoom-in on the expected central part of the source. The red `$\times$' denotes the expected position of the core, and the fact that there is no identification of the optical host galaxy. As denoted with `\xmark' in the legend, the central region of the source is not detected in any of the radio images.}
\label{fig:remnant}
\end{figure*}

\begin{figure*} [h]
\minipage{\textwidth}
        \includegraphics[width=0.46\linewidth] {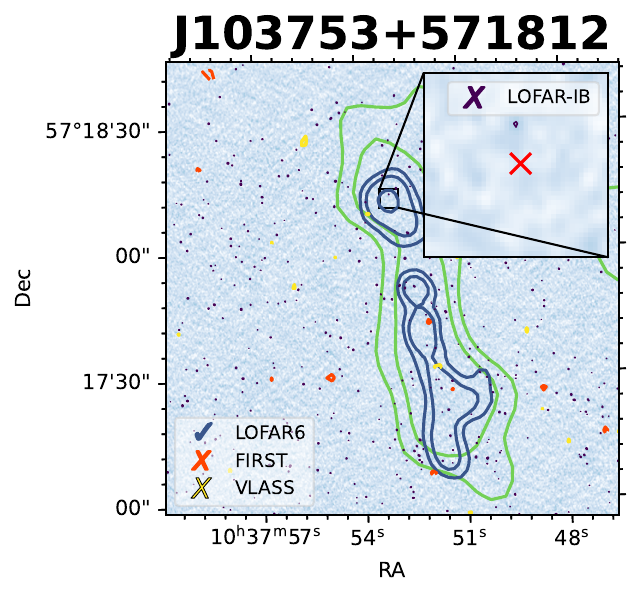}
        \includegraphics[width=0.46\linewidth] {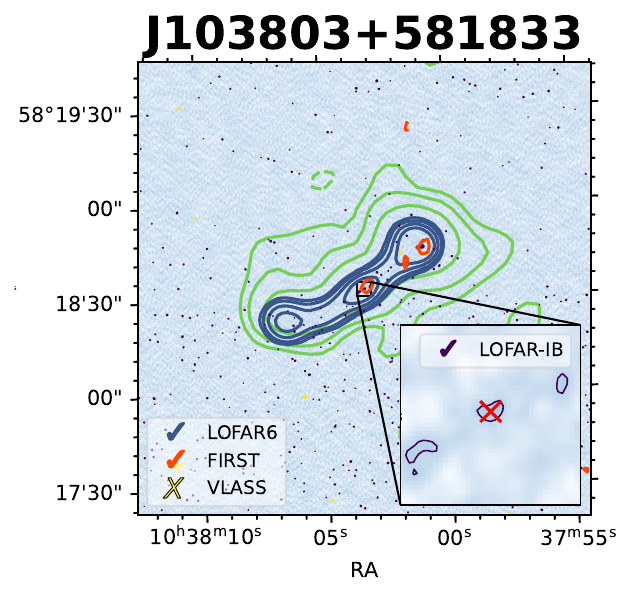}
\endminipage \hfill
\caption{Radio images of active comparison radio galaxies. 
LOFAR-IB colourmap with contours from the LOFAR18 image in green, LOFAR6 in blue, FIRST in red, VLASS in yellow, and an inset zoomed in on the central part with the LOFAR-IB contours in purple. Contour levels are the same as in Fig.~\ref{fig:restarted}. If the central part of the source is detected in radio images, it is denoted with `\cmark' in the legend. If the central part is not detected, it is noted with `\xmark'. The yellow '+' represents the position of the optical host galaxy. If the optical host galaxy is not identified, the expected position of the nuclear region is noted with red `$\times$'.}
\label{fig:COMPARISON}
\end{figure*}

\begin{figure*}[h]\ContinuedFloat
    \centering
\minipage{\textwidth}
        \includegraphics[width=0.455\linewidth] {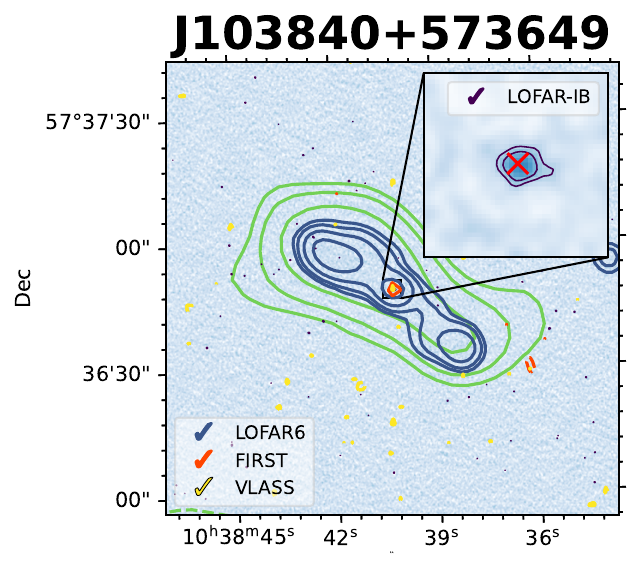}
        \includegraphics[width=0.44\linewidth] {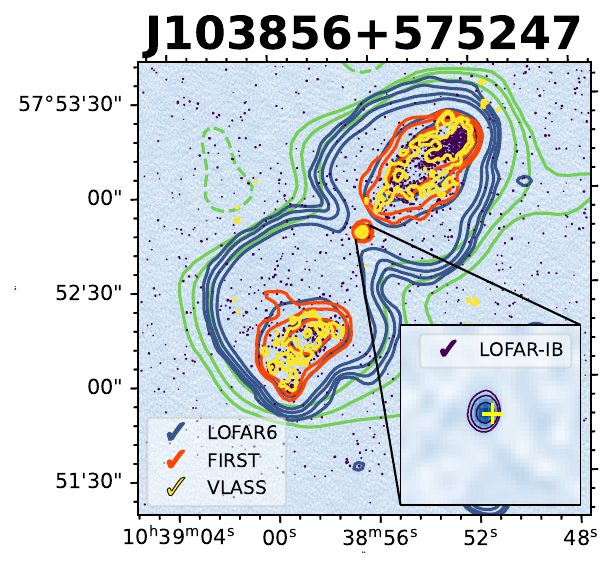}
\endminipage \hfill
\minipage{\textwidth}
        \includegraphics[width=0.455\linewidth] {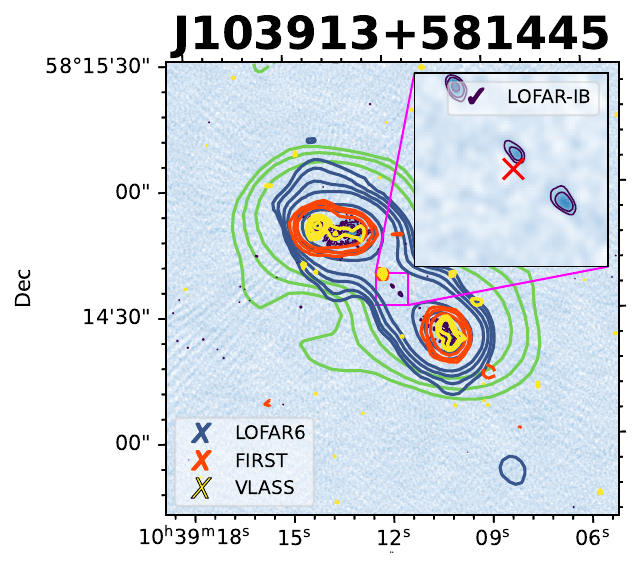}
        \includegraphics[width=0.44\linewidth] {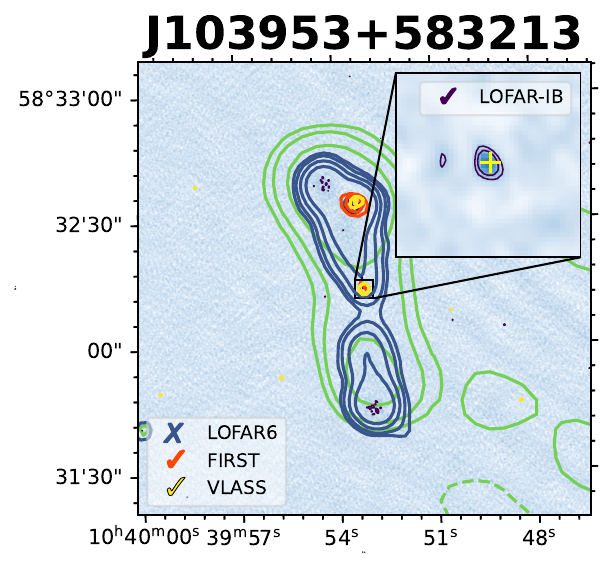}
\endminipage \hfill
\minipage{\textwidth}
        \includegraphics[width=0.455\linewidth] {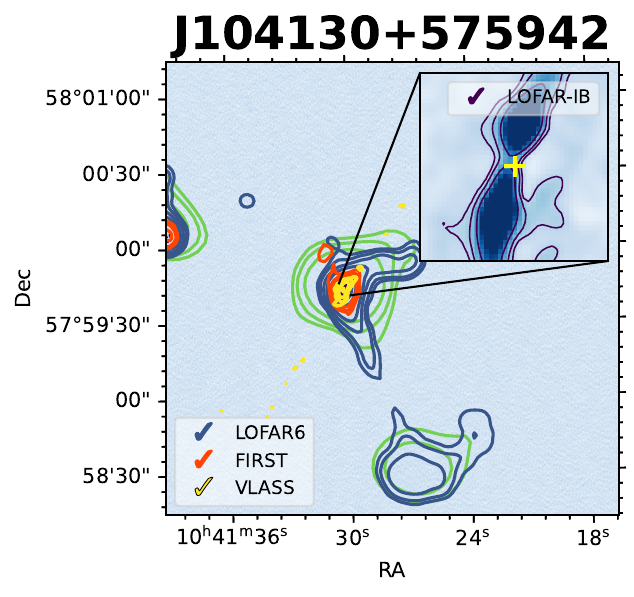} \hspace{0.45cm}
        \includegraphics[width=0.41\linewidth] {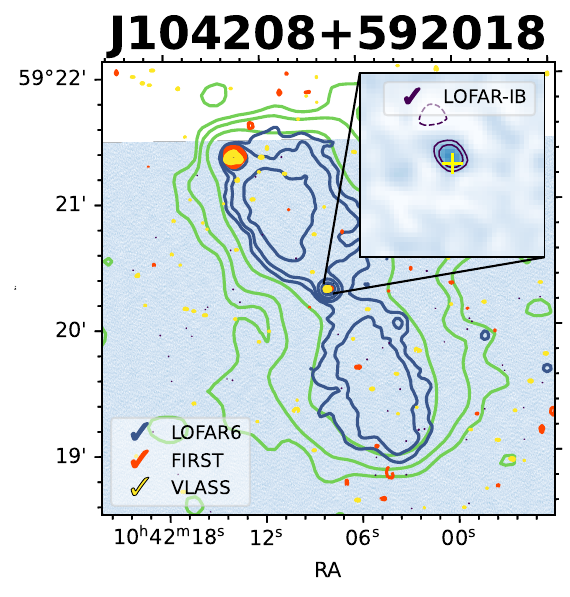}
\endminipage \hfill
\caption{Cont.}
\end{figure*}

\begin{figure*}[h]\ContinuedFloat
    \centering
\minipage{\textwidth}
        \includegraphics[width=0.46\linewidth] {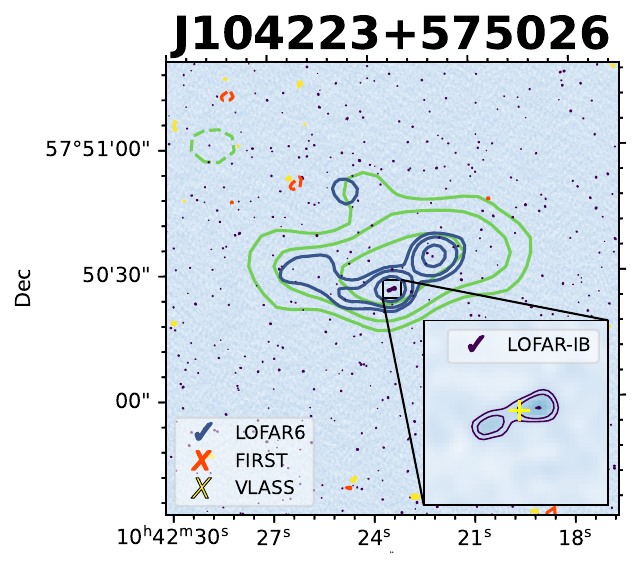}
        \includegraphics[width=0.44\linewidth] {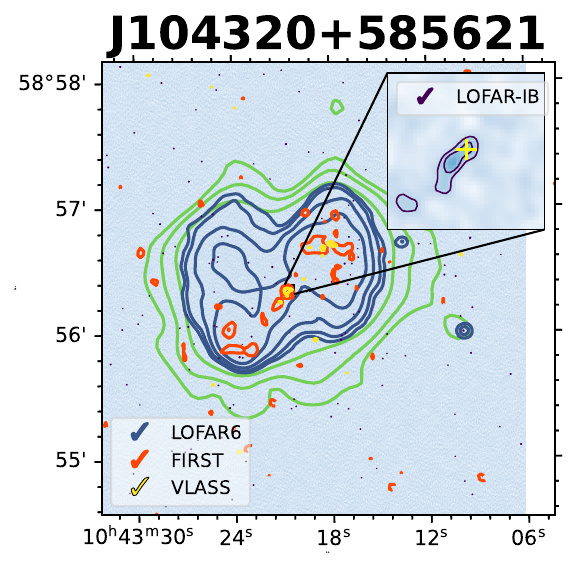}
\endminipage \hfill
\minipage{\textwidth}
\hspace{0.45cm}
        \includegraphics[width=0.425\linewidth] {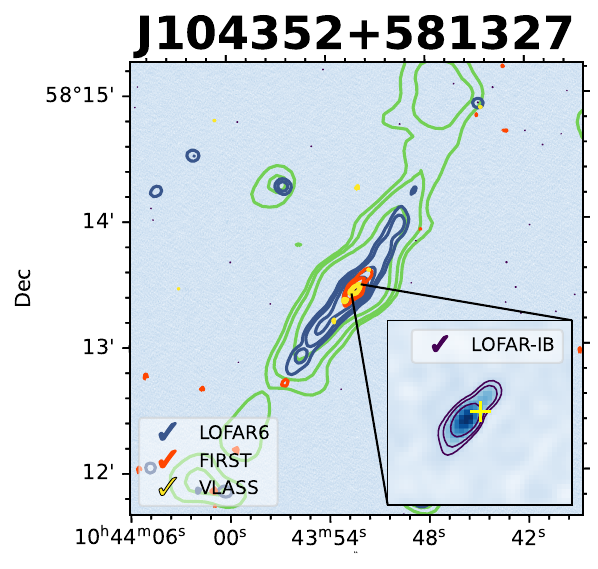}
        \includegraphics[width=0.43\linewidth] {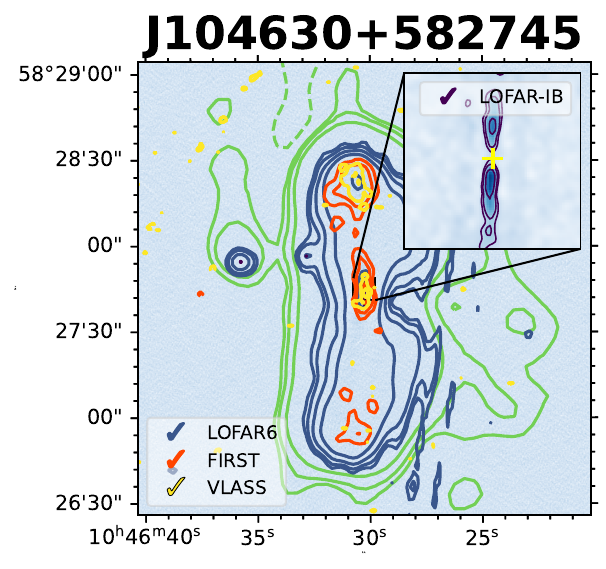}
\endminipage \hfill
\minipage{\textwidth}
        \includegraphics[width=0.465\linewidth] {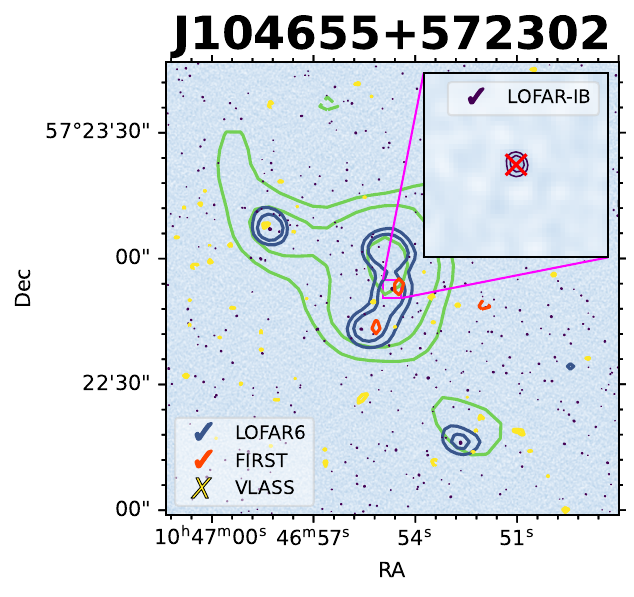}
        \includegraphics[width=0.45\linewidth] {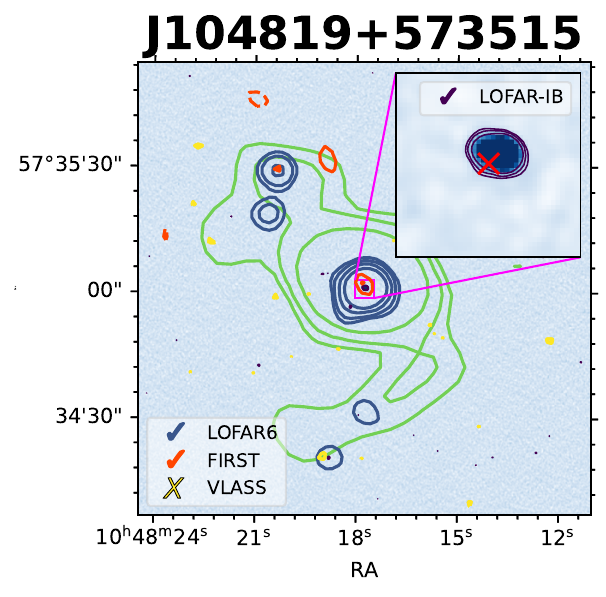}
\endminipage \hfill
\caption{Cont.}
\end{figure*}

\begin{figure*}[h]\ContinuedFloat
    \centering
\minipage{\textwidth}
        \includegraphics[width=0.47\linewidth] {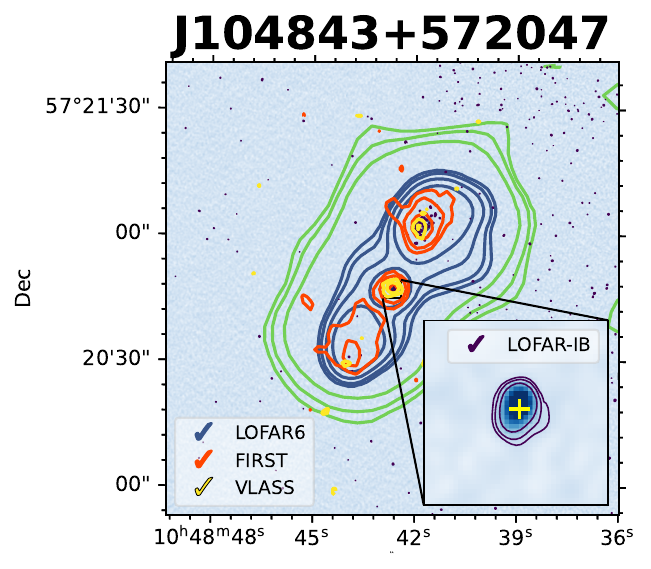}
        \includegraphics[width=0.44\linewidth] {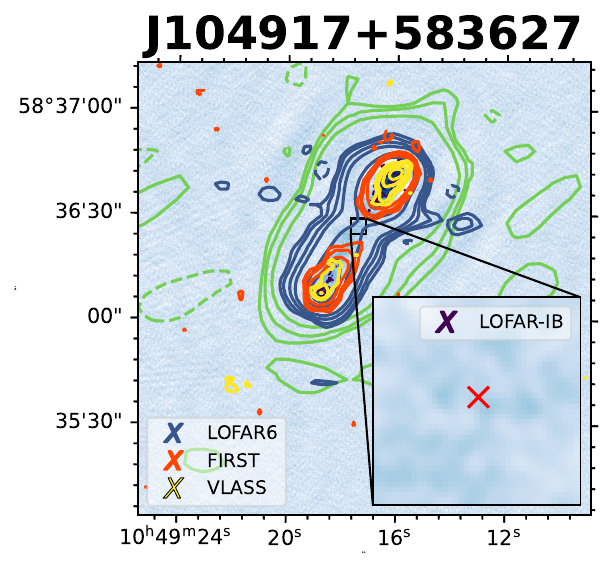}
\endminipage \hfill
\minipage{\textwidth}
        \includegraphics[width=0.47\linewidth] {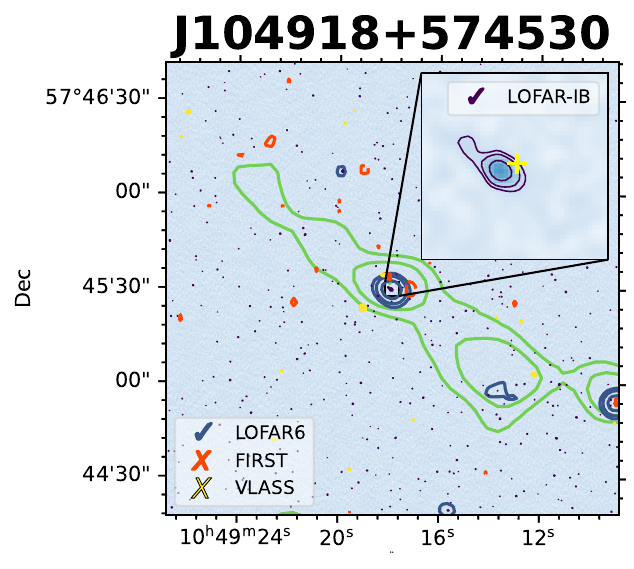}
        \includegraphics[width=0.45\linewidth] {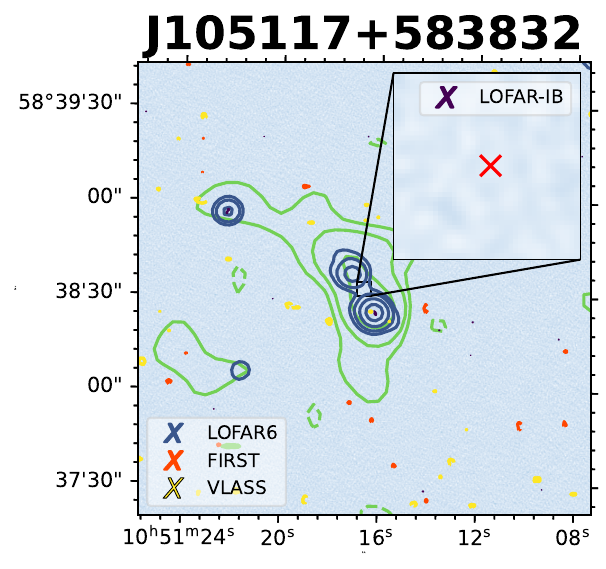}
\endminipage \hfill
\minipage{\textwidth}
        \includegraphics[width=0.47\linewidth] {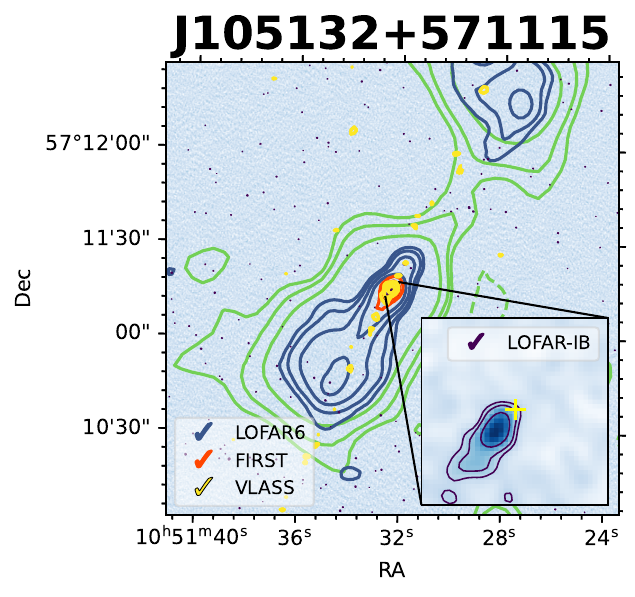}
        \includegraphics[width=0.45\linewidth] {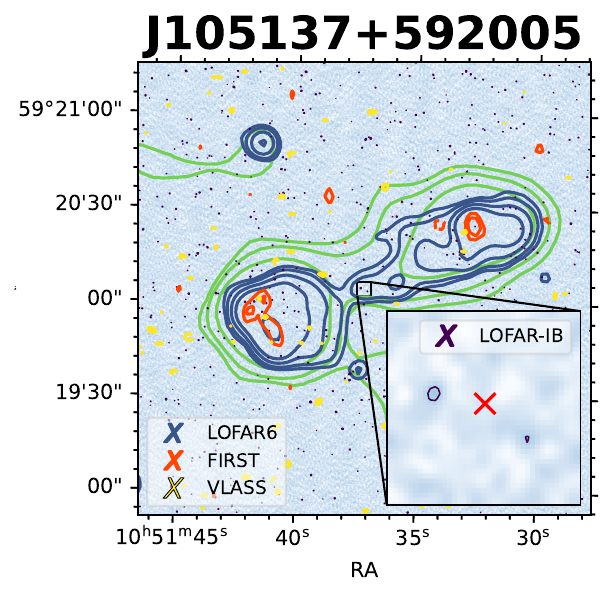}
\endminipage \hfill
\caption{Cont.}
\end{figure*}

\begin{figure*}[h]\ContinuedFloat
    \centering
\minipage{\textwidth}
        \includegraphics[width=0.47\linewidth] {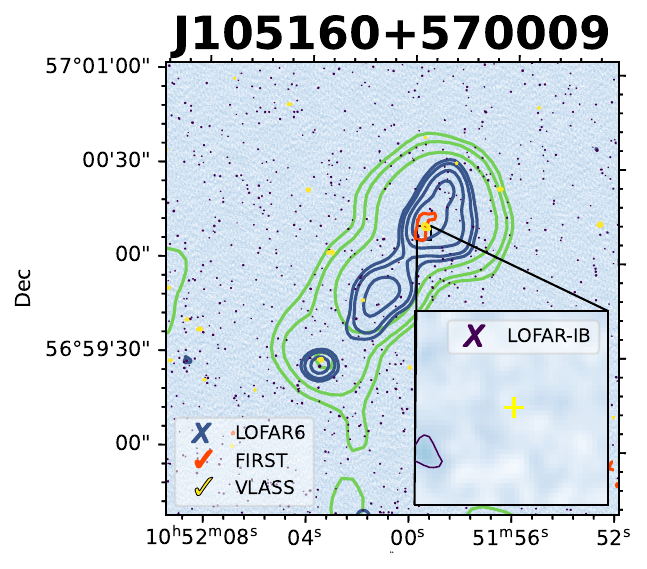}
        \includegraphics[width=0.45\linewidth] {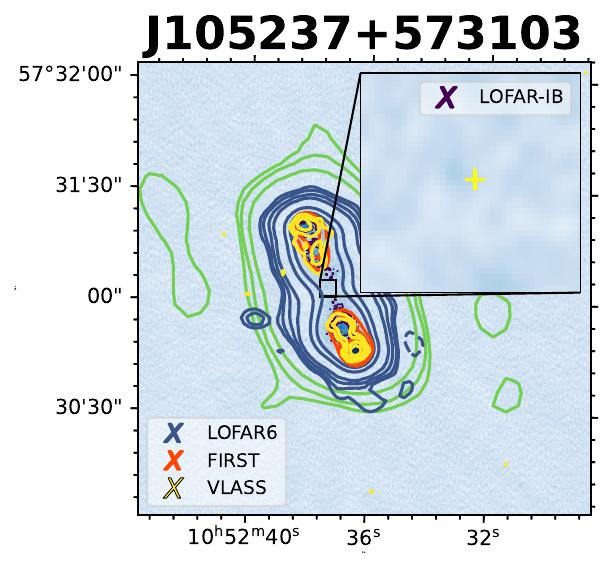}
\endminipage \hfill
\minipage{\textwidth}
        \includegraphics[width=0.47\linewidth] {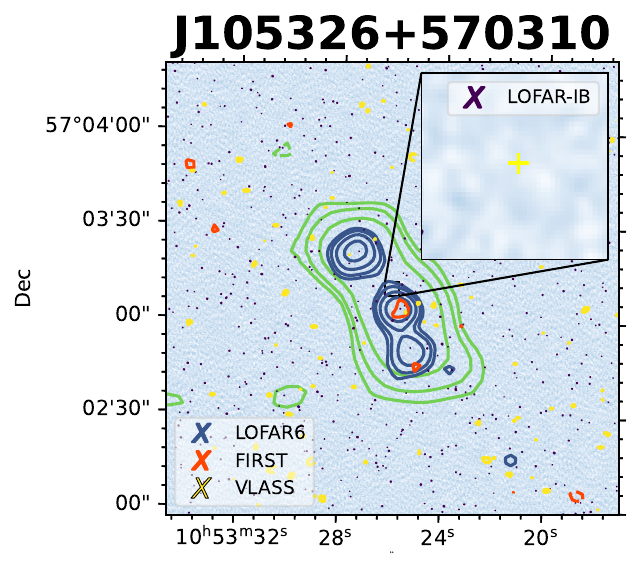}
        \includegraphics[width=0.45\linewidth] {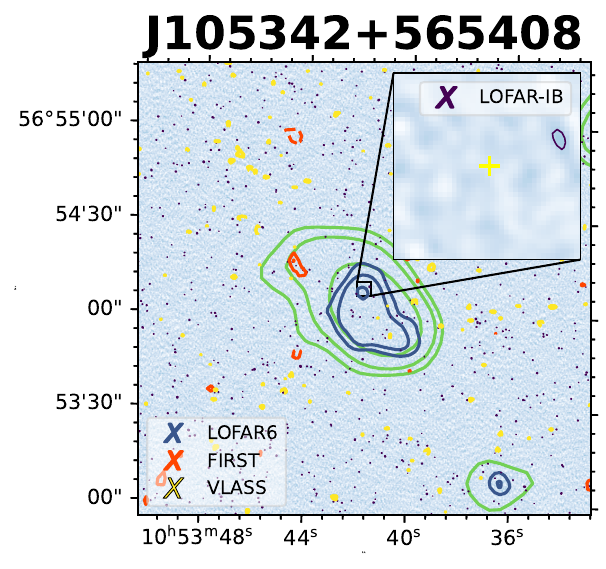}
\endminipage \hfill
\minipage{\textwidth}
        \includegraphics[width=0.47\linewidth] {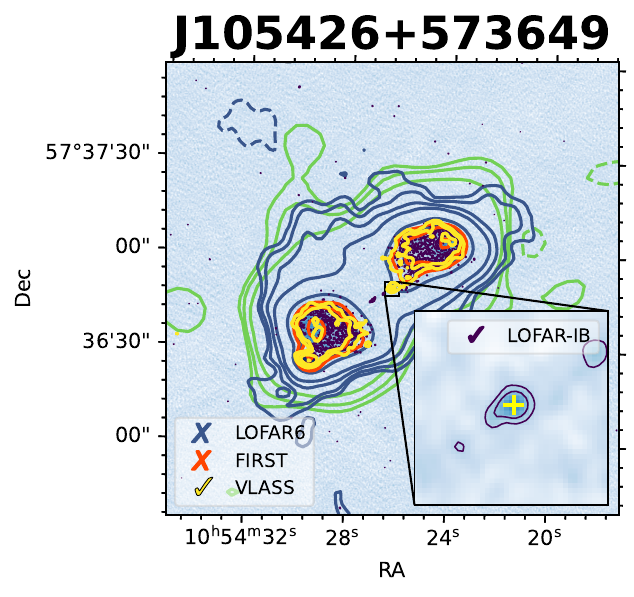}
        \includegraphics[width=0.45\linewidth] {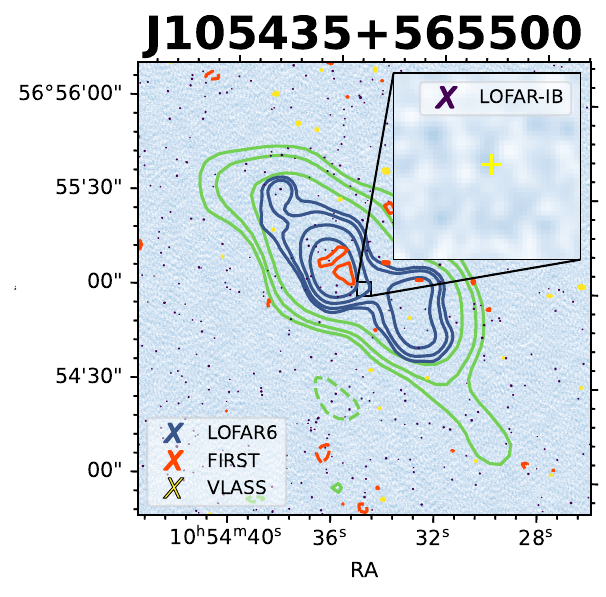}
\endminipage \hfill
\caption{Cont.}
\end{figure*}

\begin{figure*}[h]\ContinuedFloat
    \centering
\minipage{\textwidth}
        \includegraphics[width=0.47\linewidth] {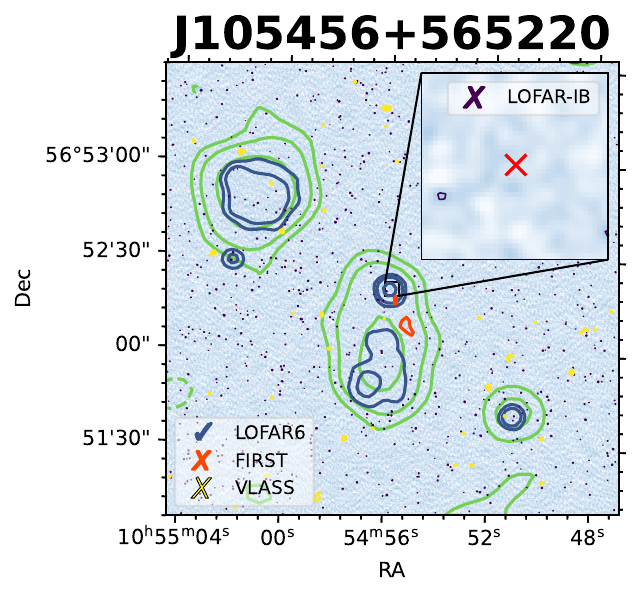}
        \includegraphics[width=0.45\linewidth] {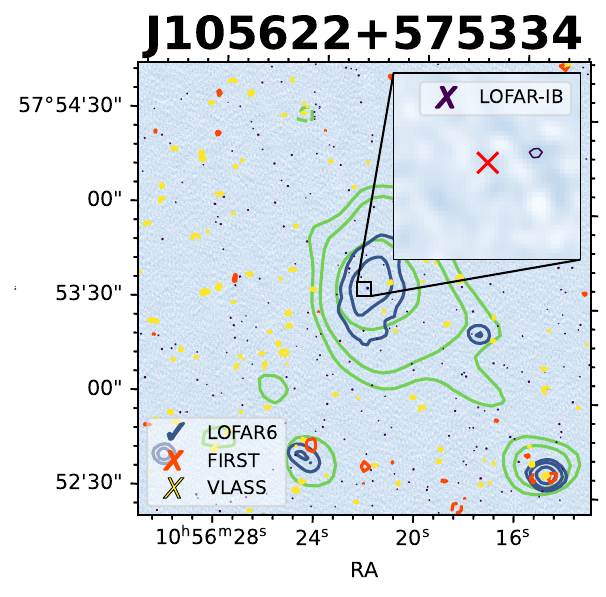}
\endminipage \hfill
\caption{Cont.}
\end{figure*}

\begin{figure*} [h]
\centering
\minipage{\textwidth}
        \includegraphics[width=0.47\linewidth] {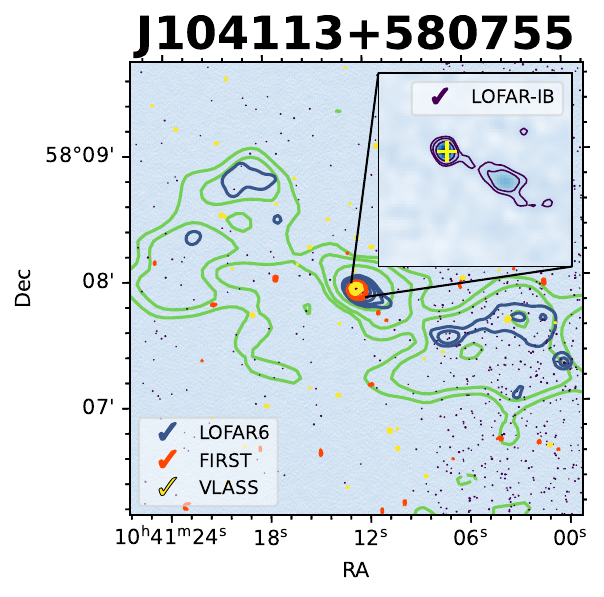}
         \includegraphics[width=0.47\linewidth] {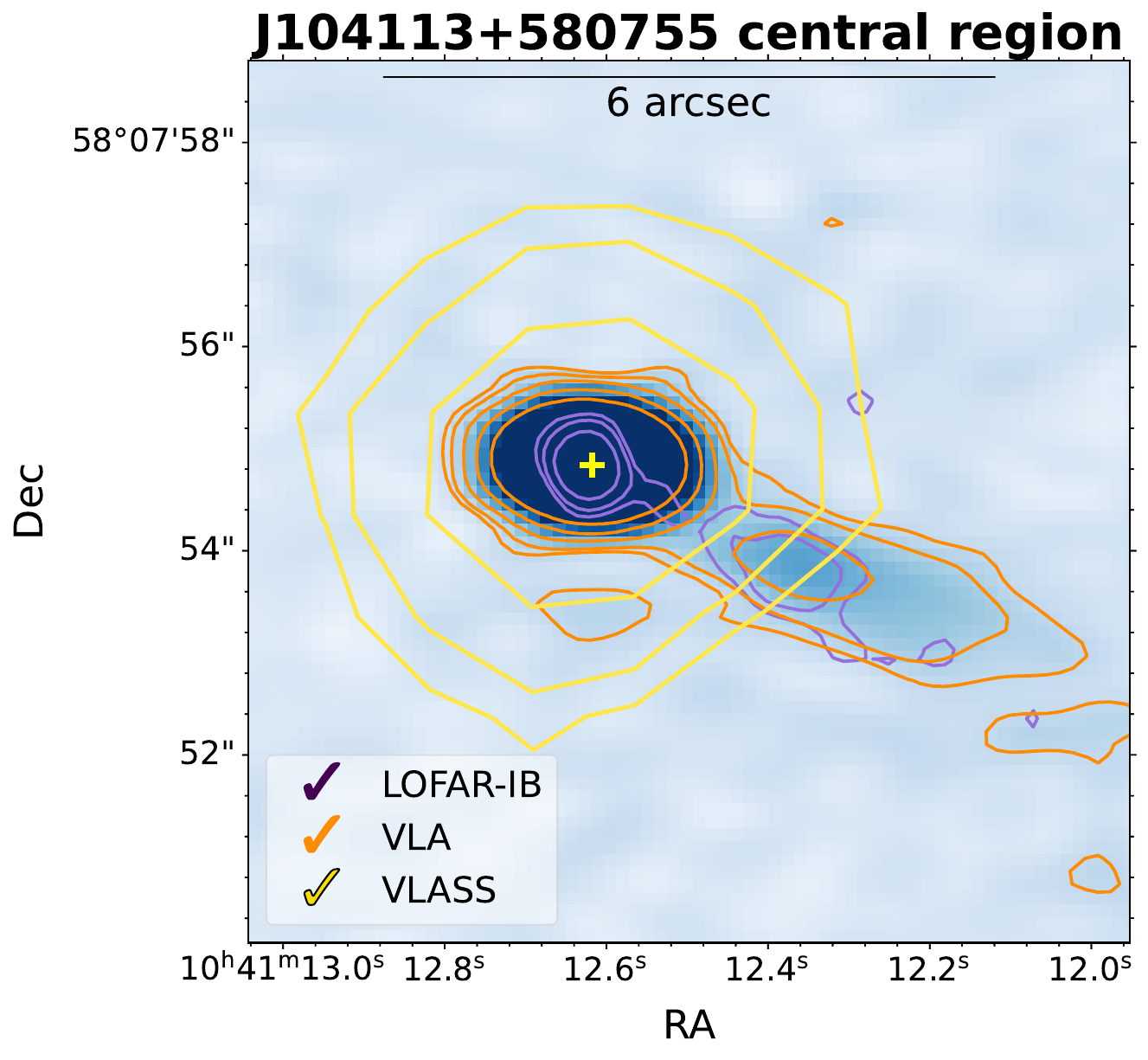}      
\endminipage \hfill
\caption{Radio images of the candidate restarted radio galaxy J104113+580755. The left panel displays the LOFAR-IB image, and the right panel shows a zoomed-in view of the central region from the dedicated VLA observation presented in Appendix~\ref{sec:Data/high-res_vla}. Contour colours and levels in the left panel are the same as in Fig.~\ref{fig:restarted}. In the right panel are LOFAR-IB contours in purple (-3, 3, 5, 10 $\times$ $\sigma$; $\sigma$ = 35 $\mu$Jy), VLA in orange (-3, 3, 5, 10, 20, 40 $\times$ $\sigma$; $\sigma$ = 12 $\mu$Jy), and VLASS in yellow (-3, 3, 5, 10 $\times$ $\sigma$; $\sigma$ = 150 $\mu$Jy). The yellow ‘+’ in the inset and in the right panel indicates the position of the optical host galaxy. The detection of the central region in the respective images is noted with `\cmark' in the legend.}
\label{fig:VLA}
\end{figure*}

\clearpage
\section{Table with radio and optical properties of the sample}

\begin{table}[h!]
\caption{Table with redshifts, total sizes of radio sources, and sizes of the central detection(s).}
\label{tab:optical}
\centering
\resizebox{\textwidth}{!}{\begin{tabular} {l c c c c c}
\hline\hline
 Source name        & redshift                                  & angular size & projected linear size & angular size & projected linear size \\
          &                                   & total [$^{\prime\prime}$] & total [kpc] & central region [$^{\prime\prime}$] & central region [kpc] \\ \hline \hline
  \multicolumn{2}{l}{Remnant}                                   &   &\\ \hline

 J104622+581427     & -                                         & 108 & 864.86    & -  & -\\\hline
  \multicolumn{2}{l}{Six restarted candidates}                  &   & \\ \hline
 J104113+580755     & 0.30894 $\pm$ 0.00006$^s$                 & 225 & 1022.85   &  3.18  & 14.45\\
 J104204+573449     & 0.4807 $\pm$ 0.0001$^s$                   & 86  & 514.11    &  1.8  & 10.76\\
 J104809+573010     & 0.31742 $\pm$ 0.00007$^s$                 & 179 & 827.88    &  0.9  & 4.17\\
 J104842+585326     & 2.1685*                                   & 70  & 590.52*    &  -  & -\\
 J104912+575014     & 0.07256 $\pm$ 0.00002$^s$                 & 106 & 146.39    &  1  & 1.38\\
 J105436+590901     & 0.8862 $\pm$ 0.0003$^s$                   & 135 & 1047.06   &  1  & 7.76\\ \hline
 \multicolumn{2}{l}{Twenty-eight active comparison sources}     &     &\\ \hline
 J103753+571812     & -                                         & 97  & 776.78     & -  & -\\ 
 J103803+581833     & 0.6467*                                   & 80  & 640.64      & 4.4  & 30.42\\ 
 J103840+573649     & 1.0657*                                   & 72  & 576.58     & 1  & 8.24\\
 J103856+575247     & 0.10078 $\pm$ 0.00003$^s$                 & 144 & 267.98     & 0.9  & 1.68\\ 
 J103913+581445     & -                                         & 85  & 690.37      & 6.1  & 49.54\\
 J103953+583213     & 1.9328 $\pm$ 0.0005$^s$                   & 75  & 629.85      & 0.8  & 6.84\\ 
 J104130+575942     & 0.31643 $\pm$ 0.00005$^s$                 & 215 & 992.44      & 8  & 37.29\\ 
 J104208+592018     & 0.5099 $\pm$ 0.0001$^s$                   & 192 & 1184.26    & 0.8  & 4.99\\ 
 J104223+575026     & 0.66 $\pm$ 0.07$^p$                       & 67  & 466.66     & 2.04  & 14.40\\ 
 J104320+585621     & 0.35294 $\pm$ 0.00005$^s$                 & 145 & 720.22     & 1.8  & 9.02\\
 J104352+581327     & 0.3898 $\pm$ 0.0002$^s$                   & 249 & 1317.21     & 1.7  & 9.08\\ 
 J104630+582745     & 0.11736 $\pm$ 0.00002$^s$                 & 134 & 283.54     & 6.5  & 13.89\\
 J104655+572302     & -                                         & 60  & 480.48      & 0.55  & 4.47\\     
 J104819+573515     & -                                         & 75  & 609.15     & 1.4  & 11.37\\ 
 J104843+572047     & 0.72 $\pm$ 0.04$^p$                       & 81  & 586.52     & 1.4  & 10.24\\
 J104917+583627     & -                                         & 85  & 680.68      & -  & -\\    
 J104918+574530     & 0.71 $\pm$ 0.05$^p$                       & 120 & 861.84      & 1.8  & 13.09\\
 J105117+583832     & 0.719*                                    & 95  & 760.76      & -  & -\\
 J105132+571115     & 0.31773 $\pm$ 0.00009$^s$                 & 155 & 718.43     & 2.2  & 10.28\\
 J105137+592005     & -                                         & 117 & 936.94     & -  & -\\  
 J105160+570009     & 0.69 $\pm$ 0.04$^p$                       & 96  & 679.97     & -  & -\\
 J105237+573103     & 0.7090 $\pm$ 0.0002$^s$                   & 78  & 560.20     & -  & -\\
 J105326+570310     & 0.52 $\pm$ 0.05$^p$                       & 73  & 454.86     & -  & -\\
 J105342+565408     & 0.24 $\pm$ 0.02$^p$                       & 60  & 226.92      & -  & -\\  
 J105426+573649     & 0.23 $\pm$ 0.06$^p$                       & 110 & 402.82      & 8.5  & 31.48\\
 J105435+565500     & 0.72 $\pm$ 0.06$^p$                       & 106 & 764.58     & -  & -\\
 J105456+565220     & -                                         & 118 & 944.94     & -  & -\\      
 J105622+575334     & 1.2657*                                   & 60  & 480.48      & -  & -\\
    \end{tabular}}
    \tablefoot{
    \small
    In Col. 1 are the source names in J2000 coordinates; Col. 2 lists redshifts of these sources determined in \cite{2020A&A...638A..34Jurlin} and \cite{2021A&A...648A...9Morganti_resolvedsi}. Those redshifts marked with $^s$ were obtained from the optical SDSS spectrum, and the ones with $^p$ are for redshifts derived from photometry. Redshifts marked with an `*' come from the LOFAR deep fields catalogue of the LH \citep{2021A&A...648A...3Kondapally,2021A&A...648A...4Duncan}. Cols. 3 -- 6 list angular and projected linear sizes of the total radio emission measured inside 3$\sigma_{\rm LOFAR18}$ contours, and the detections in the central region measured inside LOFAR-IB contours in case of one detection, while for multiple detections we measure their overall extent. Mark `-' in Cols. 5 and 6 indicates non-detection of the central region. For sources with no reported redshift, we use z = 1 to estimate linear sizes.}
\end{table}

\section{High-resolution data at 3 and 6 GHz: VLA} \label{sec:Data/high-res_vla}
In 2020, we proposed to observe the sample of restarted candidates with the \textit{Karl G. Jansky} VLA in S-band and A-configuration. However, due to the limit in the allocated time, we only observed one source, J104113+580755. This particular restarted candidate was chosen as it showed multiple components in its central region (spanning approximately 14.5 kpc) in the LOFAR-IB image, which are unresolved in the LOFAR6 image.
The observation was made with the broad-band S-band system in 8-bit mode. 
The sampling time was set to 3 seconds, and four polarisation products were recorded (RR, LL, RL, and LR). The total bandwidth, equal to $\sim$2000 MHz in the range from 2051 to 3947 MHz, was divided by default into 16 sub-bands of 128 MHz with 64 frequency channels each. 
Exposure time was set to 14.5 minutes. The observations were targeted at the candidate host galaxy. We used the quasar 3C286 as the primary flux calibrator and observed it at the beginning of the scheduling block for 11 minutes. The source J1035+5628 was used as a phase calibrator, observed before and after observing the target for the total amount of 6.25 minutes. The target restarted candidate was observed for the total amount of 14.5 minutes. This set-up of the observations is summarised in Table~\ref{tab:vla_data}. 

\begin{table} [!h]
\caption{Set-up for the VLA observations centred at 3 GHz.}
\label{tab:vla_data}
\centering                                  
\begin{tabular}{l | l}          
\hline\hline
    VLA project code            & 20B-349 \\
    Primary calibrator          & 3C~286 \\ 
    Phase calibrator            & J1035+5628 \\ 
    Bandwidth                   & 2051-3947 MHz\\
    Field of view               & 15$^{\prime}$  \\
    Configuration               & A - array \\ \hline
    Date of the observations    & 08-Jan-2021\\ 
    Observing time              & 14.5 min  \\ 
    Source observed             & J104113+580755\\ \hline \hline
\end{tabular}
\end{table}

Following the standard approach, we performed manual flagging and calibration. Half of the full band (from 2 to 3 GHz) was entirely flagged due to severe RFI in this frequency band. Phase and amplitude self-calibration was performed.
The field of view of the VLA observations ($\sim$15$^{\prime}$) was large enough to explore the presence of any compact component within the total radio source extent, which for J104113+580755 is equal to 3.75$^{\prime}$ (see Table~\ref{tab:optical} for angular and projected linear sizes of all the sources in the sample).
The angular size of the central region of the observed source detected in the LOFAR6 image is 0.1$^{\prime}$. Since this is smaller than the largest observable scale with the VLA in the A-array in S-band (0.3$^{\prime}$), we expect to recover the total flux density of the central region, if not limited by the limit in surface brightness. Therefore, we produced a uniformly weighted image and a naturally weighted one, the latter resulting in a better surface brightness sensitivity. We do not detect any other structure besides the central compact component when using robust weighting. However, we detect both the compact and the elongated component when using natural weighting.
Therefore, we use the naturally weighted image in the analysis and we show it zoomed in on the detection in the right panel in Fig.~\ref{fig:VLA}. The properties of the final images (both uniform and natural) and the peak flux densities of the detections are reported in Table~\ref{tab:vla_table}.

The final full-band image obtained with uniform weighting has a resolution of 0.64$^{\prime\prime}$ $\times$ 0.31$^{\prime\prime}$, with a position angle of -86.7 deg, and the noise of $\sim$60 $\rm \mu$Jy beam$^{-1}$. The final full-band image, produced with natural weighting, has a resolution of 1.08$^{\prime\prime}$ $\times$ 0.69$^{\prime\prime}$, with a positional angle of 86.8 deg and the noise of $\sim$12 $\rm \mu$Jy beam$^{-1}$.\\

Furthermore, we used the \textit{Karl G. Jansky} VLA image of one source from the active comparison sample (J104208+592018). This target was observed with the broad-band C-band receiver using the telescope in A-configuration on May 14, 2018. The final VLA image of the source J104208+592018 has a resolution of about 0.29$^{\prime\prime}$ $\times$ 0.23$^{\prime\prime}$, and the sensitivity of $\sim$10 $\mu$Jy beam$^{-1}$ (see Table~\ref{tab:vla_table}). The data reduction process and the final image of J104208+592018 are presented in \cite{2021A&A...653A.110Jurlin}.

For the flux density scale errors at 3 and 6 GHz, we assume a conservative value of 5\%\ (\citealt{2017ApJS..230....7Perley}).

\begin{table} [h]
\caption{VLA A-array properties of the full-band images for sources J104113+580755 (obtained using uniform and natural weighting) and J104208+592018.}
    \label{tab:vla_table}
    \centering
    \begin{tabular}{l|c|c|c} \hline \hline
Source name                         & J104113+580755                                        & J104113+580755                                        & J104208+592018                                        \\ \hline
Central frequency                   & 3 GHz                                                 & 3 GHz                                                 & 6 GHz                                                 \\ \hline
\multirow{2}{*}{Beam}               & 0.64$^{\prime\prime}$ $\times$ 0.31$^{\prime\prime}$  & 1.08$^{\prime\prime}$ $\times$ 0.69$^{\prime\prime}$  &  0.29$^{\prime\prime}$ $\times$ 0.23$^{\prime\prime}$ \\
                                    & -86.7 deg                                             & 86.8 deg                                              &  -22.2 deg                                            \\ \hline
Rms noise                           & 60  $\rm \mu$Jy beam$^{-1}$                           & 12  $\rm \mu$Jy beam$^{-1}$                           & 10  $\rm \mu$Jy beam$^{-1}$                           \\ \hline
\multirow{2}{*}{Peak flux density}  & 3.7 $\pm$ 0.2 mJy (core)                              & 4.1 $\pm$ 0.2 mJy  (core)                             &  0.48 $\pm$ 0.02 mJy                                   \\ 
                                    & <0.18 mJy (jet)                                       & 0.17 $\pm$ 0.01 mJy (jet)                                 & -    \\ \hline \hline

    \end{tabular}
\end{table}

\subsection{Detailed analysis of the restarted candidate J104113+580755} \label{sec:Results/J104113}
The source J104113+580755 has two components observed in its central region in the LOFAR-IB image. The angular size of the two detections combined is $\sim$3.18$^{\prime\prime}$, corresponding to a projected linear size of $\sim$14.45 kpc. 
Detections in the central region at 150 MHz (LOFAR-IB), 1.4 GHz (FIRST), and 3 GHz (VLASS) allowed us to derive the spectral indices of the central region presented in Table~\ref{tab:flux_si_table} and construct the spectral index plot presented in Fig.~\ref{fig:spectra}. The spectral indices calculated in this way result in a convex shape of the radio spectrum of this source, which could further indicate that the source is showing multiple epochs of the AGN activity (see also Sect.~\ref{Sec:Results/si}).\\

The two components detected in the LOFAR-IB image can be seen in the inset in Fig.~\ref{fig:VLA}.
The projected linear size of the more compact detection in the LOFAR-IB image is 4 kpc, and the size of the elongated detection is 5 kpc. The integrated flux densities of these two components in the LOFAR-IB image are 1.4 $\pm$ 0.5 mJy and 1.2 $\pm$ 0.4 mJy, respectively.

Thanks to the dedicated observations with the VLA centred at 3 GHz, at a higher resolution (1.08$^{\prime\prime}$ $\times$ 0.69$^{\prime\prime}$) and sensitivity (12 $\mu$Jy beam$^{-1}$) than VLASS, we can inspect the properties of its central region into greater detail.
In the VLA image, we detect two components coinciding with the detections in the LOFAR-IB image. The more compact detection has an angular size in the VLA image of $\sim$2.01$^{\prime\prime}$, corresponding to the linear size of 9.14 kpc, and the peak flux density of this component is 4.1 $\pm$ 0.2 mJy. The second, more elongated detection has an angular size in the VLA image of $\sim$3.51$^{\prime\prime}$, corresponding to the linear size of 15.95 kpc, and the peak flux density of this component is 0.17 $\pm$ 0.01 mJy. 

By combining the LOFAR-IB and VLA flux densities of the more compact detection, we derive its spectral index of -0.4 $\pm$ 0.1 ($\rm \alpha_{150~MHz}^{3~GHz}$). The more extended component in the west of the central component has a spectral index of 0.7 $\pm$ 0.1 ($\rm \alpha_{150~MHz}^{3~GHz}$).

The spectral analysis agrees with the compact component coinciding with the position of the optical host galaxy (see yellow `+' in the inset in Fig.~\ref{fig:restarted}), representing a core of the source and the elongated structure being a jet.
The inner structure detected in the LOFAR-IB image does not appear to be connected to the large-scale structure, which does not show clear jet structures. Therefore, the structures detected in the central region are more likely to represent a new activity cycle.

\end{document}